\documentclass[sigconf, nonacm]{acmart}

\usepackage{amssymb}   
\usepackage{amsmath}
	\usepackage{cancel}
\usepackage{mathtools}
\usepackage{graphicx} 
\usepackage{url}
\usepackage{enumitem} 

\usepackage{stmaryrd} 

\usepackage{upgreek} 
\usepackage{bbding} 

\usepackage{listings} 
\usepackage{tabularx} 
\usepackage{booktabs} 
\usepackage{multirow}
\usepackage{hhline} 
\usepackage{diagbox}
\usepackage{pgfplotstable} 

\usepackage{xcolor,colortbl}    

\usepackage{tikz} 
\usetikzlibrary{matrix}
\usetikzlibrary{calc}
\usetikzlibrary{math}
\usetikzlibrary{arrows.meta,positioning}
\usetikzlibrary{intersections, pgfplots.fillbetween}

\usepackage{easybmat} 

\usepackage{adjustbox}
\def\eox{\unskip\kern 10pt{\unitlength1pt\linethickness{.4pt}$\diamondsuit${}}} 

\usepackage{rotating}		

\usepackage{xspace} 

\usepackage{subcaption} 
\newcommand{\hide}[1]{}

\usepackage[ruled,noend,linesnumbered]{algorithm2e} 

\usepackage{setspace} 

\DontPrintSemicolon     
\SetNlSty{}{}{}                
\SetAlgoInsideSkip{smallskip}   

\SetAlFnt{\small}			
\SetAlCapFnt{\small}		
\SetAlCapNameFnt{\small}

\usepackage[capitalise]{cleveref} 

\crefname{section}{Section}{Sections}
\crefname{example}{Example}{Examples}
\crefname{figure}{Figure}{Figures}
\crefname{equation}{Equation}{Equations}
\crefname{definition}{Definition}{Definitions}

\hypersetup{                                                            
	pdfpagemode=UseNone,               
    pdfstartview=Fit,
    pdfpagelayout=SinglePage,       
    bookmarks=false,
    bookmarksnumbered=true,
    bookmarksopen=false,             
    pdftex=true,
    unicode,
    colorlinks=true,       
    linkcolor=magenta,          
    citecolor=cyan,  
    filecolor=magenta,      
    urlcolor=blue           
}

\usepackage{aliascnt}  	

\theoremstyle{acmdefinition}
\newaliascnt{corollary}{theorem}

\aliascntresetthe{corollary}

\newaliascnt{example}{theorem}
\newtheorem{example}[example]{Example}
\aliascntresetthe{example}

\newaliascnt{definition}{theorem}
\newtheorem{definition}[definition]{Definition}
\aliascntresetthe{definition}

\newaliascnt{proposition}{theorem}

\aliascntresetthe{proposition}

\newaliascnt{lemma}{theorem}

\aliascntresetthe{lemma}

\newaliascnt{conjecture}{theorem}

\aliascntresetthe{conjecture}

\newtheorem{questionW}{Question}
\newtheorem{resultW}{Result}

{\end{itshape}
}
\usepackage{tcolorbox}		
\tcbuselibrary{breakable,skins}		

\tcbset{examplestyle/.style={
		enhanced jigsaw,	
		colback=blue!08,	
		colframe=blue!08,	
		arc=2mm,
		boxrule=0pt,		
		left=0.6mm,
		right=0.6mm,
		left skip= 0mm,  
		right skip= 0mm, 
		top=0.6mm,		
		bottom=0.6mm,		
		breakable,		
		parbox = false,		
		before={\par\pagebreak[0]\vspace{0.5mm}\parindent=0pt},		
		after={\par\pagebreak[0]\vspace{0.5mm}\parindent=0pt},				
		bottomrule = 0mm,
		boxsep = 0mm,					
		topsep at break=0pt,			
		bottomsep at break=0pt,			
		pad at break=0mm,
		pad before break=1mm,		
		pad after break=0.6mm,		
		bottomrule at break=0mm,
		toprule at break=0mm,		
		}}

\usepackage{footnote}

\tcolorboxenvironment{example}{examplestyle}

\makeatletter
\DeclareRobustCommand*\uell{\mathpalette\@uell\relax}
\newcommand*\@uell[2]{
  \setbox0=\hbox{$#1\ell$}
  \setbox1=\hbox{\rotatebox{10}{$#1\ell$}}
  \dimen0=\wd0 \advance\dimen0 by -\wd1 \divide\dimen0 by 2
  \mathord{\lower 0.1ex \hbox{\kern\dimen0\unhbox1\kern\dimen0}}
}
\makeatother

\usepackage{marginnote}

\renewcommand{\epsilon}{\varepsilon} 

\usepackage{scalerel}
\usepackage{tikz}
\usetikzlibrary{svg.path}

\definecolor{orcidlogocol}{HTML}{A6CE39}
\tikzset{
  orcidlogo/.pic={
    \fill[orcidlogocol] svg{M256,128c0,70.7-57.3,128-128,128C57.3,256,0,198.7,0,128C0,57.3,57.3,0,128,0C198.7,0,256,57.3,256,128z};
    \fill[white] svg{M86.3,186.2H70.9V79.1h15.4v48.4V186.2z}
                 svg{M108.9,79.1h41.6c39.6,0,57,28.3,57,53.6c0,27.5-21.5,53.6-56.8,53.6h-41.8V79.1z M124.3,172.4h24.5c34.9,0,42.9-26.5,42.9-39.7c0-21.5-13.7-39.7-43.7-39.7h-23.7V172.4z}
                 svg{M88.7,56.8c0,5.5-4.5,10.1-10.1,10.1c-5.6,0-10.1-4.6-10.1-10.1c0-5.6,4.5-10.1,10.1-10.1C84.2,46.7,88.7,51.3,88.7,56.8z};
  }
}

\RequirePackage{etoolbox}
\DeclareRobustCommand\orcidicon[1]{\href{https://orcid.org/#1}{\mbox{\scalerel*{
\begin{tikzpicture}[yscale=-1,transform shape]
\pic{orcidlogo};
\end{tikzpicture}
}{|}}}}

\newcommand\vldbdoi{XX.XX/XXX.XX}
\newcommand\vldbpages{XXX-XXX}
\newcommand\vldbvolume{14}
\newcommand\vldbissue{1}
\newcommand\vldbyear{2026}
\newcommand\vldbauthors{\authors}
\newcommand\vldbtitle{\shorttitle} 
\newcommand\vldbavailabilityurl{URL_TO_YOUR_ARTIFACTS}
\newcommand\vldbpagestyle{plain}

\newcommand{\eat}[1]{}

\newcommand{\our}{\textsf{TACTUS}\xspace}
\newcommand{\starmie}{\textsf{Starmie}\xspace}
\newcommand{\starmieAgg}{\textsf{Starmie}$_{agg}$\xspace}
\newcommand{\liftus}{LIFTus\xspace}
\newcommand{\liftusAgg}{LIFTus$_{agg}$\xspace}
\newcommand{\santos}{\textsf{SANTOS}\xspace}
\newcommand{\sato}{\textsf{SATO}\xspace}
\newcommand{\sherlock}{\textsf{Sherlock}\xspace}
\newcommand{\dthreeL}{D$^3$L\xspace}

\newcommand{\santosSmall}{SANTOS Small\xspace}
\newcommand{\santosLarge}{SANTOS Large\xspace}
\newcommand{\TUSSmall}{TUS Small\xspace}
\newcommand{\TUSLarge}{TUS Large\xspace}
\newcommand{\wiki}{Wiki Union\xspace}
\newcommand{\WDC}{WDC\xspace}

\newcommand{\score}{\mu\xspace}
\newcommand{\scOne}{\mu_T\xspace}
\newcommand{\scTwo}{\mu_A\xspace}
\newcommand{\Tq}{T_q\xspace}
\newcommand{\Ti}{T_i\xspace}
\newcommand{\Tj}{T_j\xspace}

\newcommand{\Tipos}{T_i^+\xspace}

\newcommand{\lake}{\mathcal{T}\xspace}
\newcommand{\lakeq}{\mathcal{T}_q\xspace}

\newcommand{\ejemb}{\mathbf{e}_j\xspace}
\newcommand{\cj}{c_j\xspace}
\newcommand{\cjemb}{\mathbf{c}_j\xspace}
\newcommand{\eremb}{\mathbf{e}_r\xspace} 

\newcommand{\crq}{c_r\xspace}

\newcommand{\candset}{\mathcal{S}\xspace}
\newcommand{\batch}{\mathcal{B}\xspace}
\newcommand{\batchP}{\mathcal{B}_a\xspace}

\newcommand{\TEnc}{\mathcal{E}_T\xspace}
\newcommand{\FTenc}{\mathrm{FT}\xspace}
\newcommand{\hq}{\mathbf{T}_q\xspace}
\newcommand{\hi}{\mathbf{T}_i\xspace}
\newcommand{\hipos}{\mathbf{T}_i^+\xspace}
\newcommand{\hj}{\mathbf{T}_j\xspace}
\newcommand{\zi}{\mathbf{z}_i\xspace}
\newcommand{\batchEmb}{\mathcal{I}}
\newcommand{\Nremi}{\mathcal{N}_{\text{rem}}^{(i)}}
\newcommand{\Nhardi}{\mathcal{N}_{\text{hard}}^{(i)}\xspace}
\newcommand{\latentPi}{\mathcal{P}_{\text{latent}}^{(i)}\xspace}
\newcommand{\Tseqi}{\mathrm{Seq}_i\xspace}

\newcommand{\Cinit}{\mathcal{S}_{\text{init}}\xspace}

\newcommand{\Vj}{\mathcal{V}_j\xspace}

\def\header{\vspace{0.6mm} \noindent}

\begin{document}
\title{Efficient and Effective Table-Centric Table Union Search in Data Lakes}

\author{Yongkang Sun}
\affiliation{%
  \institution{Hong Kong Polytechnic University}
}
\email{yongkang.sun@connect.polyu.hk}

\author{Zhihao Ding}
\affiliation{%
  \institution{Hong Kong Polytechnic University}
}
\email{tommy-zh.ding@connect.polyu.hk}

\author{Huiqiang Wang}
\affiliation{%
  \institution{Hong Kong Polytechnic University}
}
\email{huiqiang.wang@connect.polyu.hk}

\author{Reynold Cheng}
\affiliation{%
  \institution{The University of Hong Kong}
}
\email{ckcheng@cs.hku.hk}

\author{Jieming Shi}
\affiliation{%
  \institution{Hong Kong Polytechnic University}
}
\email{jieming.shi@polyu.edu.hk}

\begin{abstract} 

In data lakes, information on the same subject is often fragmented across multiple tables. Table union search aims to find the top-$k$ tables that can be unioned with a query table to extend it with more rows, without relying on metadata or ground-truth labels.
Existing methods are mainly {\em column-centric}: they focus on modeling column unionability scores using column embeddings, which are then used throughout the search process for column matching, filtering, and aggregation. However, this overlooks holistic table-level semantics, which may result in suboptimal rankings and inefficiencies.

We introduce \our, a novel {\em table-centric} method for table union search. Unlike prior work that searches from columns to tables, we search in a table-first way and examine columns only in the final step.
During offline processing, we directly generate table embeddings for holistic, table-level unionability scoring by designing {\em table-level representation techniques}, including positive table pair construction to simulate unionable tables, two-pronged negative table sampling to avoid latent positives and mine hard negatives to enhance representation quality, and attentive table encoding for effective embeddings.
During online search, we first develop a {\em table-centric adaptive candidate retrieval} method that efficiently selects a compact, high-quality candidate pool by leveraging the distribution of table-level unionability scores induced by table embeddings. 
We then inspect columns only within this compact candidate set and design a {\em dual-evidence reranking} technique that integrates table-level and column-level scores to refine the final top-$k$ results.
Extensive experiments on real-world datasets show that \our significantly improves result quality while being much faster than existing methods in both offline and online processing, often by an order of magnitude.

\end{abstract} 

\maketitle

\pagestyle{\vldbpagestyle}
\begingroup\small\noindent\raggedright\textbf{PVLDB Reference Format:}\\
\vldbauthors. \vldbtitle. PVLDB, \vldbvolume(\vldbissue): \vldbpages, \vldbyear.
\endgroup
\begingroup
\renewcommand\thefootnote{}\footnote{\noindent
This work is licensed under the Creative Commons BY-NC-ND 4.0 International License. Visit \url{https://creativecommons.org/licenses/by-nc-nd/4.0/} to view a copy of this license. For any use beyond those covered by this license, obtain permission by emailing \href{mailto:info@vldb.org}{info@vldb.org}. Copyright is held by the owner/author(s). Publication rights licensed to the VLDB Endowment. \\
\raggedright Proceedings of the VLDB Endowment, Vol. \vldbvolume, No. \vldbissue\ %
ISSN 2150-8097. \\
\href{https://doi.org/\vldbdoi}{doi:\vldbdoi} \\
}\addtocounter{footnote}{-1}\endgroup

\ifdefempty{\vldbavailabilityurl}{}{
\vspace{.3cm}
\begingroup\small\noindent\raggedright\textbf{PVLDB Artifact Availability:}\\
The source code, data, and/or other artifacts have been made available at \url{https://github.com/MyriadVerse/TACTUS}.
\endgroup
}

\section{Introduction}\label{sec:intro}

Tabular data on the web, from government, scientific, and commercial sources, are increasingly available~\cite{google_dataset_search,castelo2021auctus,fernandez2018aurum,limaye2010annotating}. Data on the same subject is often fragmented across multiple tables, published by different organizations~\cite{lehmberg2017stitching,nargesian2018table,ling2013synthesizing}, making it crucial to discover relevant tables in data lakes to construct richer datasets in various applications~\cite{miao2023watchog,zhang2020finding,miller2018open}. However, web tables often lack metadata (e.g., headers or schemas)~\cite{adelfio2013schema,farid2016clams,nargesian2019data,zhang2020finding}, rendering conventional search inadequate~\cite{google_dataset_search}. This motivates table search that relies on table content rather than curated metadata~\cite{nargesian2018table,sarma2012finding}.

In this paper, we focus on the task of Table Union Search~\cite{nargesian2018table,starmie,liftus}: given a query table $\Tq$, the goal is to find the top-$k$ tables in a data lake $\lake$ that can be unioned with $\Tq$ to extend it with additional rows. 
This task does not use table metadata or ground-truth labels, making it practical.  Two tables are considered unionable if they describe the same subject and their columns are semantically aligned (e.g., tables on funded research projects published by different government divisions~\cite{nargesian2018table}, or tables on video games published at different times~\cite{lehmberg2017stitching}). Unioning such tables with different rows yields a larger, semantically consistent table.

\begin{figure}[!t]
    \centering
    \includegraphics[width=1.01\linewidth]{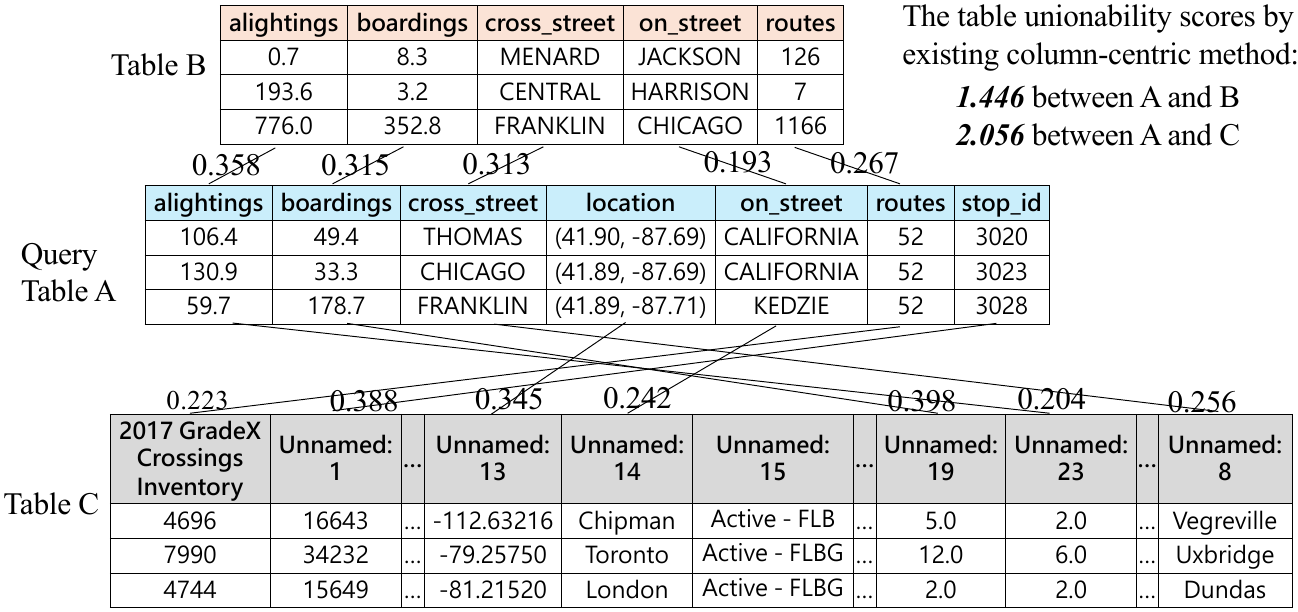}
    \caption{
    Tables~A and~B are unionable as they share the same subject (bus ridership) and have semantically aligned columns. Table~C, on a different subject (railway maintenance), is non-unionable with table~A; unioning them leads to inconsistent table content. The links between columns indicate the highly matched column pairs with column unionability scores from~\cite{starmie}, which are aggregated into table unionability scores (shown top-right); it ranks non-unionable table~C (2.056) above unionable table~B (1.446) for query table~A.
    }
    \label{fig:union-example}
    \vspace{-5mm}
\end{figure}

\begin{example}
Consider the tables in \cref{fig:union-example}. We include column names to make the example easy to follow, while such metadata is often missing in practice and we do not assume its availability.
Query table~A and table~B are unionable because they (i) concern the same subject (bus ridership data) and (ii) have semantically aligned columns. Intuitively, they can be unioned by merging their distinct rows to produce a larger table with consistent meaning. 
In contrast, table~C concerns a different subject (Canada railway maintenance data) and is non-unionable with table~A, even though some columns may incidentally share similar values (e.g., numerical columns).
Forcing to union tables~A and~C would mix rows about bus ridership and railway maintenance, resulting in a table with undesired inconsistent content.

\end{example}

Finding unionable tables has been widely studied, as reviewed in \cref{sec:sota}~\cite{ling2013synthesizing,lehmberg2017stitching,starmie,liftus,nargesian2018table,d3l,santos}. Notably, \citet{nargesian2018table} formalized the table union search problem and evaluated various features for measuring unionability. Methods differ in how they compute unionability scores. With advances in representation learning and pretrained language models (PLMs), recent methods such as \starmie~\cite{starmie} and \liftus~\cite{liftus} learn column embeddings to score column unionability, and then aggregate these scores to derive table unionability scores for ranking. This alleviates the need for manual feature engineering and explicit cell value matching. 

State-of-the-art methods~\cite{starmie,liftus} mainly adopt a {\em column-centric} approach, focusing on modeling column unionability scores:
(i) they first learn {\em column embeddings} for every column in the data lake $\lake$; 
(ii) during online search, for a query table $\Tq$, they compute column unionability scores by similarities between column embeddings of $\Tq$ and those in $\lake$, to filter and retrieve candidate tables; and (iii) for each candidate table $\Ti$, they then perform column matching between $\Tq$ and $\Ti$ and aggregate the scores of highly-matched column pairs into a table unionability score for ranking to get the top-$k$ results. 

However, this column-centric approach suffers from the following issues.
First, it proceeds early to column-level matching, overlooking the overall table-level semantics, which may lead to counterintuitive rankings as illustrated below.

\begin{example}\label{example:bias}
In \cref{fig:union-example}, \starmie computes pairwise column unionability scores and applies maximum-weight bipartite matching to align columns between tables~A and~B.
The five links between tables~A and~B represent matched column pairs with column unionability scores by \starmie, and summing them up  yields a table unionability score of 1.446. For tables~A and~C, \starmie finds seven matched columns with a higher aggregated score of 2.056, causing table~C to be ranked above table~B, despite being non-unionable with table~A. 
This misranking occurs since table~C has columns with high column unionability scores to columns in table~A (e.g., `Unnamed: 19' and `Unnamed: 13' in table~C score 0.398 and 0.345 with `boardings' and `location' in table~A). Aggregating them let table~C to outrank the truly unionable table~B.
\end{example}

We believe that effective table union search should consider both holistic table-level semantics and  column matches, while existing methods focus mainly on the columns and largely overlook table-level semantics. Further, for efficiency, prior methods use column-based filtering: retrieve any table with at least one column highly matched to any query column in $\Tq$. As $\Tq$ has multiple columns, each matching columns in many tables, this yields a large candidate set with non-unionable tables, causing unnecessary overhead.

Motivated by these findings, we propose \our, a \underline{TA}ble-\underline{C}entric method for efficient and accurate \underline{T}able \underline{U}nion \underline{S}earch.
Our approach performs search in a {\em table-to-column} manner, in contrast to the existing column-to-table paradigm.

\our first ensures that the overall table-level semantics of a table $\Ti$ are promising for union with $\Tq$, and then refines it by integrating both table-level and column-level patterns. 
Specifically, we generate a {\em table embedding} $\hi$ for each table $\Ti \in \lake$ to capture table-level unionability.
During online search, for a query $\Tq$, we first use table embeddings to compute a table-level unionability score $\scOne(\Tq,\Ti)$ and design {\em table-centric adaptive candidate retrieval} to obtain a compact candidate set $\candset$.
For example, in \cref{fig:union-example}, \our assigns a high score $\scOne(A,B)=0.805$ and a low score $\scOne(A,C)=0.233$, retaining table~B for further inspection and pruning non-unionable table~C early.
Since we filter using a single embedding per table, unlike prior methods that use multiple column embeddings from $\Tq$, our candidate retrieval  yields a compact, distribution-aware pool $\candset$ with controllable size: for each query, it adaptively selects candidates based on the   distribution of $\scOne$ scores to get a concise, high-coverage $\candset$ with size typically close to $k$.
Over $\candset$, we develop {\em dual-evidence reranking}: we combine the table-level unionability score $\scOne(\Tq,\Ti)$ with an extra column alignment score $\scTwo(\Tq,\Ti)$ to compute the final table unionability score $\score(\Tq,\Ti)$, and rank candidates by $\score(\Tq,\Ti)$ for the top-$k$.  
Note that $\scOne$ from table embeddings is used for retrieval and ranking, while $\scTwo$ is computed only for candidates in the last step.

During offline stage, we obtain a table embedding $\hi$ for each $\Ti \in \lake$ to capture table-level unionability, in contrast to prior work~\cite{starmie,liftus} that learns column embeddings for all columns in $\lake$.
Since ground-truth labels are unavailable, we develop self-supervised table-level representation techniques to obtain table embeddings, ensuring that unionable tables are close while non-unionable tables are distant in the embedding space.
Specifically, we introduce {\em positive table pair construction} to simulate unionable table pairs, and a {\em two-pronged negative table sampling} strategy that excludes latent positives and mines hard negatives to enhance table embeddings for unionability estimation. We develop an {\em attentive table encoder} with tailored serialization and multi-head attention to produce table embeddings, trained with a table-level contrastive objective.
While contrastive learning is popular across domains, the key differences among methods lie in how they instantiate its core components. We design these dedicated techniques for table pairs to get  effective table embeddings, as detailed in~\cref{sec:contrastive}, in contrast to prior methods~\cite{starmie,liftus} that focus on column pairs.

We conduct extensive experiments and show that \our consistently achieves superior result quality while being significantly faster than existing methods in both offline and online stages, often by an order of magnitude.
Our contributions are as follows:
\vspace{-\topsep+1pt}
\begin{itemize}[leftmargin=*]
    \item We propose \our, a novel table-centric framework for efficient and accurate table union search in data lakes.
    \item We design an efficient table-to-column search process, introducing table-centric adaptive candidate retrieval and dual-evidence reranking techniques.
    \item To obtain effective table embeddings, we develop positive table pair construction, two-pronged negative table sampling, and attentive table encoding methods.
    \item Extensive experiments demonstrate that \our consistently achieves state-of-the-art effectiveness and efficiency.
\end{itemize}

\section{Preliminaries}
\label{sec:preliminaries}

\subsection{Table Union Search Problem}

A data lake $\lake$ is a collection of $n$ tables. 
Each table $T \in \lake$ consists of multiple columns and rows.

The top-$k$ table union search problem~\cite{nargesian2018table,starmie,liftus} in \cref{def:topktus} aims to identify a subset $\lakeq \subseteq \lake$ of size $k$ that contains the top-$k$ tables $T$ with the highest table unionability scores $\score(\Tq, T)$ to a query table $\Tq$. 

Note that there is no universal definition of the table unionability score $\score(\cdot,\cdot)$; different methods compute it based on their own design choices, as discussed in \cref{sec:sota}. In our approach, we first develop the table-level unionability score $\scOne(\Tq,T)$ from table embeddings for candidate filtering, and then combine the table-level $\scOne(\Tq,T)$ with a column-level alignment score $\scTwo(\Tq,T)$ to get our final unionability score $\score(\Tq,T)$ for reranking. 
 
Moreover, there are two problem settings to emphasize~\cite{nargesian2018table,starmie,liftus}. (i) In real-world data lakes, metadata such as column headers or schemas are often missing or unreliable~\cite{santos,nargesian2018table,srinivas2023lakebench,lehmberg2016large}. Thus, table union search {\em does not use any table metadata} and relies solely on table content for unionability assessment.
(ii) The table union search is generally formulated as an {\em unsupervised} problem, since ground-truth unionability labels for table pairs are typically unavailable. In the experiments, benchmark datasets with such labels are used solely for effectiveness evaluation.

\vspace{-0.5mm}
\begin{definition}[Top-$k$ Table Union Search]
\label{def:topktus} 
Given a data lake $\lake$ and a query table $\Tq$, the  table union search problem is to find a subset $\lakeq \subseteq \lake$ with $|\lakeq| = k$ such that for every $T \in \lakeq$ and every $T' \in \lake \setminus \lakeq$, $\score(\Tq, T) \ge \score(\Tq, T')$, where  $\score(\cdot,\cdot)$ is the table unionability score.
\end{definition}

\subsection{Related Work}
\label{sec:sota}

\noindent
\textbf{Table Union Search.} Early work relied on keywords, string/set-based similarity, and schema information to identify related tables~\cite{cafarella2009data, sarma2012finding, harmouch2021relational, LiLL08, PimplikarS12}. \citet{ling2013synthesizing} defined table stitching as the task of unioning tables with identical schemas within a site, while \citet{lehmberg2017stitching} merged web tables using metadata such as column headers. \citet{sarma2012finding} framed the problem of finding unionable web tables as the entity complement problem, introducing entity and schema consistency measures for search. These works underscore the need to find unionable tables but rely on table metadata, which is often unavailable in practice~\cite{santos, nargesian2018table, srinivas2023lakebench}.

\citet{nargesian2018table} formalized table union search without metadata, defining its own column unionability score based on statistical value overlap, ontology annotations, and language features, and computing table unionability through optimal one-to-one column alignment.
\dthreeL~\cite{d3l} measures column similarity using multiple features, including column names, and employs hashing to retrieve candidate tables.
\citet{santos} extend the table union search framework~\cite{nargesian2018table} by requiring an intent column in the query table, and propose \santos, which incorporates relationship semantics between columns and knowledge bases for search.
\starmie~\cite{starmie} is the first to apply contrastive learning with PLMs for table union search, generating column embeddings that capture contextual semantics and quantify column unionability via cosine similarity. \liftus~\cite{liftus} further improves column representations by incorporating both linguistic and non-linguistic features.

These methods~\cite{starmie,liftus,nargesian2018table}, especially \starmie and \liftus, primarily adopt a column-centric approach for table union search. They focus on first modeling column unionability scores, which are then used throughout the search process and aggregated to get table unionability scores.
For example, the state-of-the-art \starmie method generates column embeddings for all columns in the data lake $\lake$.
During online search, it retrieves candidate tables that contain at least one column highly unionable with any column in the query table $\Tq$, based on column unionability scores.
It then computes the table unionability score between $\Tq$ and each candidate $\Ti$ using maximum-weight bipartite matching over their column unionability scores, and returns the top-$k$ unionable tables.

As illustrated in \cref{sec:intro} (\cref{fig:union-example}), column-centric approaches focus on column details too early and overlook the overall table-level semantics, which can result in irrelevant unions—such as combining a bus ridership table with a railway maintenance table that are clearly not unionable. An experimental study~\cite{abs-2505-21329} also shows  the importance of table-level semantics.   

We argue that effective table union search should first capture holistic table-level semantics and then refine results with fine-grained column details.
Accordingly, we propose our table-centric \our method, which searches from tables to columns, rather than from columns to tables in prior work.  We produce one embedding per table to compute a table-level unionability score $\scOne$, enabling efficient candidate retrieval and early pruning of tables on unrelated subjects. Only over the candidate set, we inspect columns, get a column-level score $\scTwo$, and combine it with $\scOne$ into our final table unionability score $\score$ for top-$k$ ranking.
By leveraging table embeddings, \our operates on a smaller candidate set and improves both effectiveness and efficiency.

Besides, a recent work~\cite{DUST} introduces a new problem of finding diverse unionable tuples for  query tuples, and develops the DUST method based on embedding and clustering techniques. Another related but different problem is join discovery, to find tables that can be joined to expand columns horizontally~\cite{dong2021efficient,dong2023deepjoin,Snoopy,santos2022sketch,zhu2019josie,MATE,Ember,Omnimatch}. 
For example, MATE~\cite{MATE} efficiently discovers multi-column (n-ary) key joins at scale using a space-efficient hash-based super key index, while the Ember system~\cite{Ember} abstracts and automates keyless joins by constructing an index with task-specific embeddings. These studies highlight the importance and ongoing research in tabular data discovery.

\header 
\textbf{Representation Learning of Tables.}
Representation learning has been widely adopted for various tabular data tasks. 
For example, {\sherlock}~\cite{sherlock} and \sato~\cite{sato} generate embeddings to predict the semantic types of columns by designing rich features and tailored models.
Prior work~\cite{starmie,liftus} has adapted these representations for table union search evaluation, and we also include them in the experiments.
Other studies leverage PLMs  for table understanding~\cite{deng2022turl}, entity matching~\cite{cappuzzo2020creating,0001LSDT20,LiLSWHT21,tu2023unicorn}, question answering~\cite{IidaTMI21}, schema matching~\cite{koutras2021valentine,tu2023unicorn}, table transformation~\cite{LiHYWC23,li2023auto}, and table overlap estimation~\cite{PugnaloniZPLNS25}. We build on this line of research, but design dedicated embedding, scoring, and search techniques for table union search.
There are also studies on general table representation learning that focus on effective pre-training~\cite{jia2023getpt,chen2023hytrel}, but these pursue objectives different from table union search, which aims to efficiently find accurate sets of unionable tables.
Recently, large language models (LLMs) have shown great potential in data management~\cite{liu2024magneto,Pneuma,kayali2024chorus,BIRDIE}. 
For example, \citet{liu2024magneto} proposed Magneto, a schema matching method that combines the strengths of small and large language models to achieve state-of-the-art performance.
\citet{kayali2024chorus} applied LLMs to data discovery and exploration, demonstrating a promising direction toward unifying various tasks under foundation models.   
These developments also motivate us to explore LLMs for table union search in future work.

\begin{figure}[t!]
    \centering
    \includegraphics[width=1\linewidth]{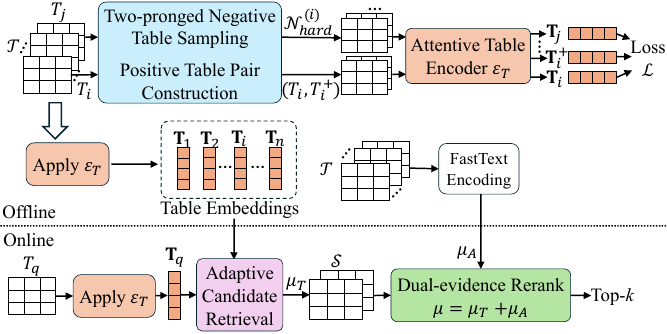}
        \vspace{1pt}
    \caption{The Overview of \our}
    \label{fig:overview}
    \vspace{-2mm} 
\end{figure} 

\section{Solution overview}\label{sec:overview} 
Here we present a high-level overview of our method in~\cref{fig:overview}, while detailed designs are elaborated in subsequent sections.

In the offline stage, given a data lake $\lake$, we generate a \emph{table embedding} $\hi$ for each table $\Ti\in\lake$. 
Since ground-truth unionability labels are unavailable, we design methods to construct positive and negative table pairs for each table $\Ti$ in~\cref{sec:contrastive-fw}.
Specifically, we first construct a positive table $\Tipos$ for $\Ti$ by jointly sampling rows and columns, ensuring that $\Ti$ and $\Tipos$ share similar semantics while also containing distinct content, thereby closely simulating real-world unionable scenarios. Next, we apply {\em two-pronged negative table sampling}: we exclude tables that are likely unionable with $\Ti$ (i.e., latent positives), and mine hard negative tables $\Nhardi$ for $\Ti$ that are challenging to distinguish from true positives. This approach enhances the quality of table embeddings for unionability estimation.
Next, we develop an {\em attentive table encoder} in \cref{sec:contrastive-att} that generates table embeddings using dedicated serialization and multi-head attention aggregation. The table encoder $\TEnc$ is trained to pull positive table pairs $(\Ti, \Tipos)$ closer and push negative tables of $\Ti$ further apart. Applying the trained table encoder $\TEnc$ to all tables in $\lake$ yields their table embeddings $\batchEmb = \{\hi \mid \Ti \in \lake\}$.
Our table embeddings are designed to capture table-level unionability semantics, so their similarity directly serves as the table-level unionability score $\scOne(\Ti, \Tj)$ between tables $\Ti$ and $\Tj$.

In the online stage, given a query table $\Tq$, we first apply {\em table-centric adaptive candidate retrieval} (\cref{sec:retrieval}), which uses the table-level unionability score $\scOne(\Tq, \Ti)$ from table embeddings to efficiently identify promising candidate tables.
Specifically, we perform a single nearest neighbor search over the indexed table embeddings $\batchEmb$ using the query table embedding $\hq$ to retrieve initial candidates, and then adaptively select a concise candidate pool $\candset$ based on the distribution of table-level unionability scores, typically yielding a small, high-quality set with size close to $k$.
Over the candidate set $\candset$, we further refine the ranking by integrating both table-level and column-level semantics. Specifically, we employ a {\em dual-evidence reranking} method (\cref{sec:rerank}) that combines the table-level unionability score $\scOne(\Tq, \Ti)$ and a column alignment score $\scTwo(\Tq, \Ti)$ to compute the final table unionability score $\score(\Tq, \Ti)$. 
The column alignment score $\scTwo(\Tq, \Ti)$ uses column embeddings computed offline via FastText encoding on basic features as described in \cref{sec:rerank}.
We then return the top-$k$ tables with the highest $\score(\Tq, \Ti)$ from $\candset$ as the result.

\section{Table-Centric Offline Processing}
\label{sec:contrastive}
We aim to produce a table embedding $\hi$ for each table $\Ti$ in the data lake $\lake$, such that unionable tables are close in the embedding space while non-unionable tables are well separated.

A straightforward approach is to aggregate column embeddings from existing methods~\cite{starmie, liftus} to form a table embedding $\hi$. However, as shown in~\cref{tab:aggregating-column} of the experiments, this is suboptimal: column embeddings are trained for column-level unionability, and simple aggregation is insufficient to capture holistic table semantics. Thus, we propose a direct table-level representation method to produce table embeddings that reflect table-level unionability.

\subsection{Table-Level Unionability Modeling}
\label{sec:contrastive-fw}

Given two tables, $\Ti$ and $\Tj$, if they are unionable, their embeddings $\hi$ and $\hj$ should be close in the embedding space; otherwise, they should be far apart. However, in the unsupervised problem setting, ground-truth table unionability labels are unavailable for training table embeddings. 
Contrastive learning~\cite{simclr} is a widely used technique for learning data representations without supervision.
Applying contrastive learning requires the following components: (i) constructing positive pairs of similar instances, (ii) sampling negative instances that are dissimilar, and (iii) designing an encoder to map instances to embeddings, together with a suitable contrastive loss to train over positive and negative samples.

While contrastive learning is a general framework, the key differences between methods hinge on how they implement the components. These design choices are critical for the quality of the resulting representations. 
\starmie~\cite{starmie} pioneers this direction for table union search by applying contrastive learning at the {\em column} level: constructing positive and negative column pairs, and performing column encoding to preserve column unionability via a column-pair-based contrastive loss.
 
In contrast, our approach differs in all components by designing techniques at the {\em table} level. Specifically, we introduce: (i) a positive table pair construction technique to simulate table unionability in the absence of labels; (ii) a two-pronged negative table sampling strategy that excludes latent positives and mines hard negatives that are challenging to distinguish from positives, enabling the table embeddings to better capture unionability patterns; and (iii) an attentive table encoder with dedicated serialization and multi-head attention aggregation in \cref{sec:contrastive-att}, trained with a table-level contrastive loss on table pairs.

\header 
\textbf{Positive Table Pair Construction.} Given a training batch $\batch$ of $N$ tables, we construct a positive table $\Tipos$ for each table $\Ti$ in $\batch$, forming a positive table pair $(\Ti, \Tipos)$. 
Prior work~\cite{starmie, liftus} constructs positive column pair  and explores various sampling strategies at the cell, column, and row levels for different datasets.  

Instead, we construct positive table pairs as follows.
(1) For each table $\Ti$, we randomly select a proportion of rows and a proportion of columns to retain, and sample these to obtain $\Tipos$. The sampled rows are shuffled to increase the diversity of the positive table while preserving semantic consistency.
(2) Note that $\Tipos$ is a subset of the original table $\Ti$, i.e., every value in $\Tipos$ appears in $\Ti$. However, real unionable tables often contain not only overlapping but also distinct values. To better simulate this, we also apply row sampling to the original table $\Ti$, so that $\Ti$ and $\Tipos$ have distinct content.
This approach simulates real-world scenarios where unionable tables exhibit variations in both rows and columns while preserving overall semantics. For notational simplicity, we continue to use $\Ti$ to denote this row-sampled version hereafter.

Given a batch $\batch$ of $N$ tables, we construct $N$ positive pairs $(\Ti, \Tipos)$, resulting in $2N$ tables: the row-sampled originals $\batch$ and their corresponding positives $\batch^+ = \{\Tipos\}_{i=1}^N$. Together, these form the augmented batch $\batchP = \batch \cup \batch^+$. 
Hereafter, for any table $\Ti \in \batchP$ where $i=1,\dots,2N$, we denote its positive table as $\Tipos \in \batchP$.

\header 
\textbf{Two-pronged Negative Table Sampling.} 
For each table $\Ti$ in batch $\batchP$, we need to identify its negative table samples, i.e., tables that are likely non-unionable with $\Ti$.

A naive approach is to treat all tables in $\batchP \setminus \{\Ti,\Tipos\}$ as negatives. However, this has two issues.
First, it may include false negatives: some tables in $\batchP \setminus \{\Ti,\Tipos\}$ may actually be unionable with $\Ti$ (i.e., latent positives), which is unknown a priori. Penalizing these as negative tables during training degrades representation quality.
Second, some tables in $\batchP \setminus \{\Ti,\Tipos\}$ may be easy negatives that are clearly non-unionable with $\Ti$, offering little learning signal and wasting computation. Including such easy negatives can dilute the effect of informative samples, so training should focus on hard negatives that are more challenging to distinguish from positives~\cite{robinsoncontrastive,liu2024magneto}.

We thus propose a two-pronged negative table sampling technique: (i) exclude latent positives—tables with a high likelihood of being unionable with the query table, and (ii) select hard negatives—remaining tables that are difficult to distinguish from the query table. Both steps are performed dynamically per batch using the current table embeddings.

First, for latent positive exclusion, within each training batch $\batchP$, we use the current table embeddings to identify and exclude tables that are likely unionable with $\Ti$ from its negative set.
Specifically, for tables $\Ti$ and $\Tj$ in $\batchP$, their table embeddings $\hi$ and $\hj$ are designed such that their cosine similarity reflects their table-level unionability score $\scOne(\Ti,\Tj)$, as will be detailed in \cref{eq:single-loss}. 
If $\scOne(\Ti,\Tj)$ exceeds a threshold, we treat $\Tj$ as a latent positive of $\Ti$ in batch $\batchP$. For a table $\Ti$ in $\batchP$, its latent positive set $\latentPi$ consists of tables $\Tj$ (excluding $\Ti$ and its predefined positive $\Tipos$) with score $\scOne$ to $\Ti$ exceeds a threshold $\gamma$, set to 0.9, in \cref{eq:latent-set}:
\begin{equation}\label{eq:latent-set}
\latentPi = \left\{\, \Tj \in \batchP \ \middle| \ \Tj \neq \Ti,\ \Tj \neq \Tipos,\ \scOne(\Ti,\Tj) > \gamma \,\right\}.
\end{equation}

\begin{figure}
    \centering    \includegraphics[width=0.6\linewidth]{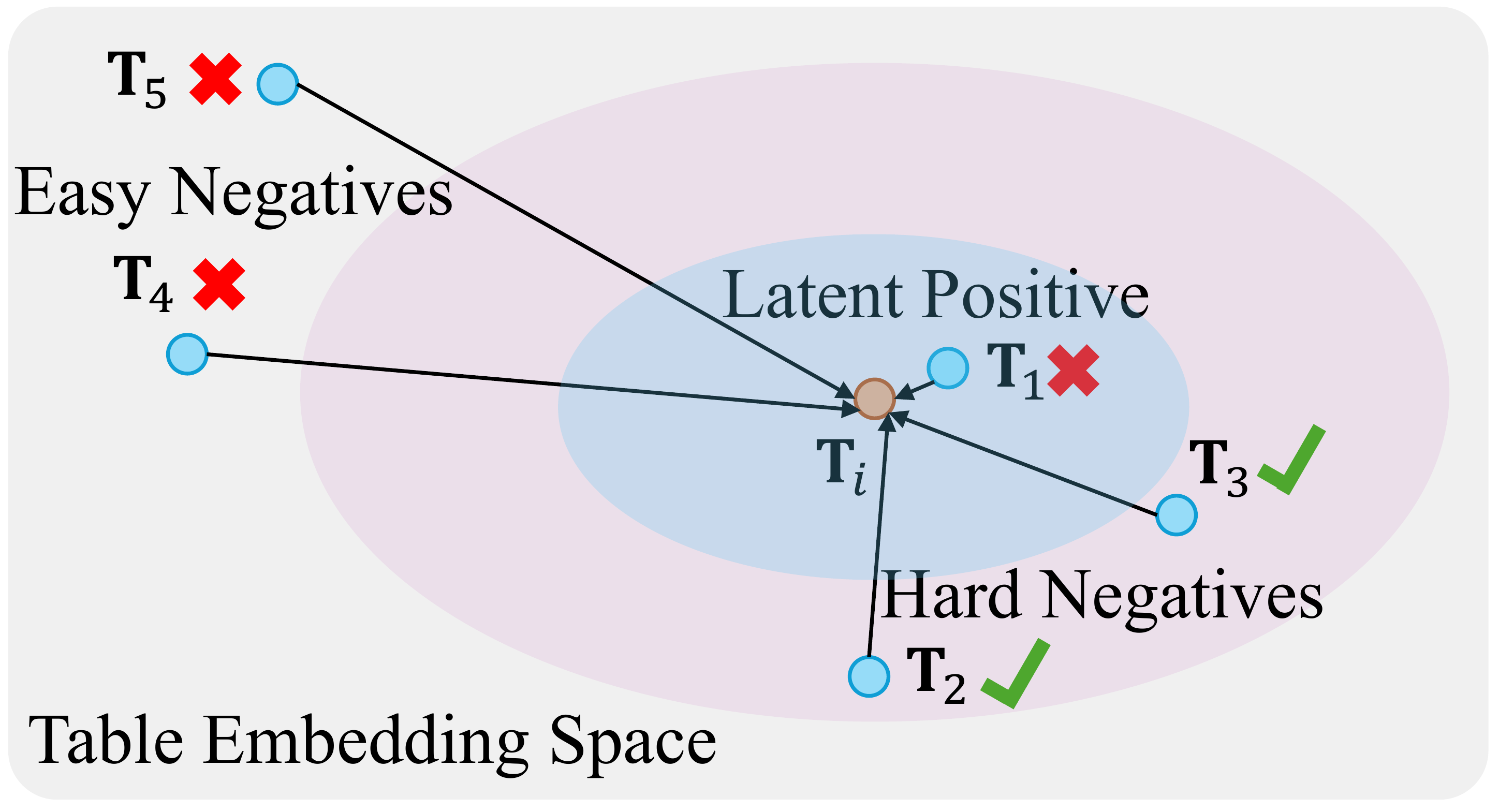}
    \caption{Illustration of Two-pronged Negative Sampling}
    \label{fig:embedding-space} 
    \vspace{-3mm}
\end{figure}

We exclude these latent positives in $\latentPi$ from being selected as negative samples for table $\Ti$ in batch $\batchP$, resulting in the remaining negative candidates $\Nremi$ for $\Ti$,  in \cref{eq:hard-neg} below.

Second, for hard negative table sampling, we aim to identify the most informative negative tables $\Nhardi$ for $\Ti$ from $\Nremi$, as shown in the second line of \cref{eq:hard-neg}. 
Intuitively, tables that are relatively similar to $\Ti$ but not unionable are harder to distinguish and thus provide stronger training signals.
Specifically, we rank tables in $\Nremi$ by their score $\scOne(\Ti,\Tj)$, and retain the top half as hard negatives $\Nhardi$. Since latent positives with high scores have already been excluded from $\Nremi$, it is reasonable to treat the remaining tables with relatively high scores as hard negatives.
The lower half of tables in $\Nremi$ with low scores are considered easy negatives and are thus excluded from training $\hi$ in the batch.
\begin{equation}\label{eq:hard-neg}
\begin{aligned}
    \Nremi &= \batchP \setminus \big(\latentPi \cup \{\Ti,\Tipos\}\big), \\
\Nhardi &= \underset{\Tj \in \Nremi}{\operatorname{arg\,top-}\tfrac{|\Nremi|}{2}}\Big\{\scOne(\Ti,\Tj)\Big\},
\end{aligned}
\end{equation}
where  $\operatorname{arg\,top-}|\Nremi|/2$ returns the $|\Nremi|/2$ elements of $\Nremi$.

As illustrated in \cref{fig:embedding-space}, the embedding space can be conceptually divided into three regions: 
the innermost region near $\Ti$ contains latent positives (e.g., $T_1$), which are excluded from negative sampling due to their high likelihood of being unionable with $\Ti$; 
the outermost region far from $\Ti$ contains easy negatives (e.g., $T_4$ and $T_5$), which are clearly non-unionable and thus provide little training signal; 
the middle region contains hard negatives (e.g., $T_2$ and $T_3$), which are non-unionable but similar to $\Ti$, making them challenging and informative for training. 
These hard negatives are retained, i.e., $\Nhardi = \{T_2, T_3\}$.

By employing two-pronged negative sampling, each table $\Ti$ in $\batchP$ is with a negative set $\Nhardi$ whose size may vary. This allows to focus training on the informative negatives, thereby enhancing the effectiveness of the table embeddings for unionability estimation.

\header
\textbf{Training Objective.}
Our objective adopts cosine similarity between table embeddings to quantify table-level unionability $\scOne(\Ti,\Tj)$. 
For table $\Ti$   with its positive table $\Tipos$ and hard negatives $\Nhardi$, we use the InfoNCE loss for the pair $(\Ti,\Tipos)$  in \cref{eq:single-loss}, where $\tau$ is a temperature parameter typically set to $0.07$~\cite{starmie,liftus}:
\begin{equation}\label{eq:single-loss}
\ell(\Ti,\Tipos) = - \log \frac{\exp\!\big(\scOne(\Ti,\Tipos)/\tau\big)}
{\sum_{\Tj \in \Nhardi} \exp\!\big(\scOne(\Ti,\Tj)/\tau\big)},
\end{equation}

Then the batch loss is computed by averaging the single-pair loss over all $2N$ tables in $\batchP$. Minimizing this loss encourages each $\hi$ to be close to its positive table $\hipos$, while simultaneously pushing $\hi$ away from its hard negatives in $\Nhardi$.
\begin{equation}\label{eq:full-loss}
\mathcal{L} = \frac{1}{2N} \sum_{i=1}^{2N} \ell(\Ti, \Tipos).
\end{equation}

\subsection{Attentive Table Encoding}
\label{sec:contrastive-att} 
We develop a table encoder $\TEnc$ that generates table embeddings $\hi$ for tables $\Ti$. Recent work on various tasks~\cite{suhara2022annotating,miao2023watchog,deng2022turl,bussotti2023generation,dong2023deepjoin} leverages PLMs to serialize a table into a text sequence to obtain contextualized token embeddings. We are not reinventing this pipeline. Instead, we focus on designing techniques to adapt and enhance the serialization and encoding steps to support our table-level unionability representation modeling in \cref{sec:contrastive-fw}.

\header 
\textbf{Serialization.}
Given a table $\Ti$, we serialize it into a token sequence suitable for input to a PLM such as BERT~\cite{bert}. A straightforward way is to concatenate all cell values within each column and join columns using special delimiter tokens (e.g., \texttt{[CLS]}). However, due to the token budget limit of PLMs, this approach does not prioritize informative tokens and may allocate budget to less relevant content~\cite{liu2024magneto,dong2023deepjoin}.

Existing table union search methods~\cite{starmie,liftus} use various heuristics to rank and sample cell values for serialization: \starmie applies TF-IDF or alphabetical selection, while \liftus ranks values by string length. In contrast, we adopt priority sampling~\cite{priority-sample}, which has proven effective in schema matching~\cite{liu2024magneto}, to select informative tokens for serialization. Priority sampling favors values with higher weights while introducing controlled randomness. In our approach, word frequency (TF) serves as the weight, which we find effective. 
Specifically, for each column $\cj$ in table $\Ti$, we tokenize and compute the frequency of each unique value $w$ in the column as TF($w$). The weight of $w$ in $\cj$ is then calculated as $\text{TF}(w)/h(w)$, where $h(w)$ is a random function in $[0.8,1]$. All unique values in each column $c_j$ are then ranked by their weights in descending order. For all columns in $\Ti$, we select the top-ranked values for serialization until the token limit is reached.

\header
\noindent{\bf Attentive Table Encoder.} 
Given a table $\Ti$ with $m$ columns, let $\Tseqi = \texttt{[CLS]}w_1^1 w_1^2 \dots \texttt{[CLS]} w_m^1 w_m^2 \dots$ denote its token sequence, where $w_j^i$ is the $i$-th ranked token in column $c_j$. This sequence is fed into BERT, which uses Transformer layers with self-attention~\cite{vaswani2017attention} to obtain contextualized embeddings. The special delimiter token \texttt{[CLS]} marks the start of each column, and the hidden state at each \texttt{[CLS]} position in the final layer of the PLM is used as the embedding $\cjemb$ for column $c_j$, capturing its semantics in the context of the entire table. This yields $m$ column embeddings $\{\cjemb\}_{j=1}^m$ for table $\Ti$.  

Unlike existing methods that use column embeddings directly for column-centric table union search, we require a unified table embedding $\hi$ that captures the overall unionability of $\Ti$ as modeled in  \cref{sec:contrastive-fw}. To achieve this, we integrate the embeddings $\cjemb$ of table $\Ti$ into an intermediate embedding vector $\zi$ using a multi-head attention mechanism~\cite{vaswani2017attention}, as shown in \cref{eq:aggregate}. We then refine $\zi$ via a multilayer perceptron (MLP) projection to obtain the final table embedding $\hi$ for table $\Ti$, as in \cref{eq:mlp}. The MLP consists of two linear layers with a ReLU activation function~\cite{relu} between the layers. 

\begin{figure}[!t]
\centering
\includegraphics[width=0.94\linewidth]{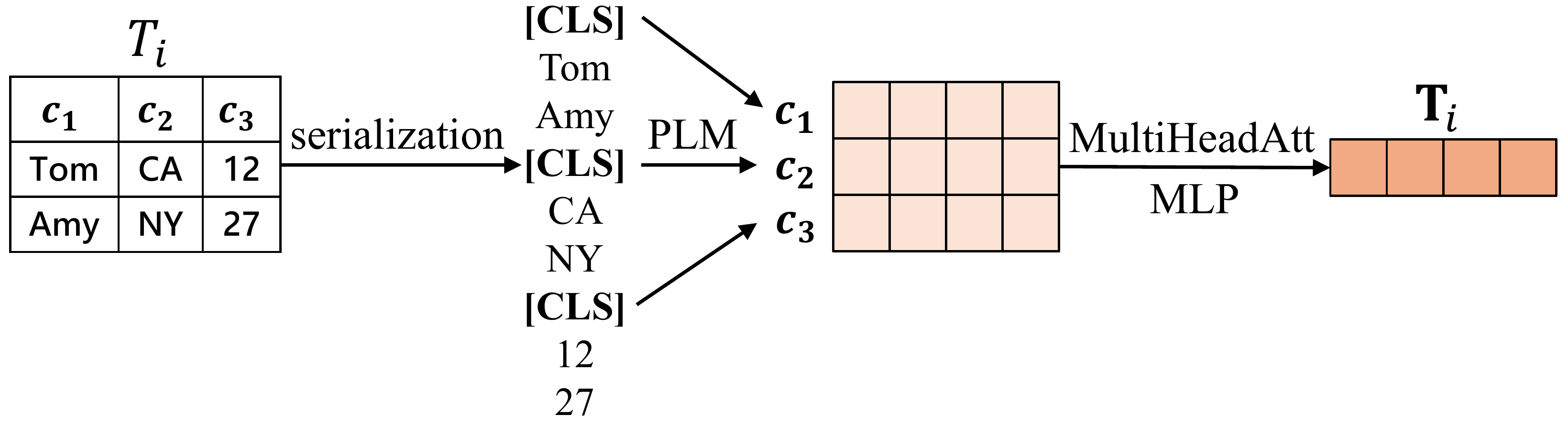}
\caption{Illustration of the  Attentive Table Encoder}
\label{fig:attention-aggr}
\vspace{-2mm}
\end{figure}

\cref{fig:attention-aggr} illustrates the attentive table encoder with the dedicated serialization and aggregation steps as designed above.
\begin{equation}\label{eq:aggregate}
\zi = \text{MultiHeadAtt}(\{\cjemb\}_{j=1}^m; Q, K, V),
\end{equation}
\begin{equation}\label{eq:mlp}
\hi = \text{MLP}(\zi),
\end{equation}
where $Q$ is a learnable parameter, and $K, V$ are $\{\cjemb\}_{j=1}^m$.

\begin{algorithm}[t]
\caption{\our Training}
\label{alg:contrastive-learning}
\KwIn{
A data lake   $\lake$}
\KwOut{Trained table encoder $\TEnc$}
\For{$\text{ep} = 1$ \KwTo $n_{\text{epoch}}$}{
    Randomly partition $\lake$ into batches\;
    \ForEach{batch $\batch$}{
            $\batchP \gets \emptyset$\tcp*[l]{Augmented batch}  
        \ForEach{$\Ti \in \batch$}{
            Construct positive table pair $(\Ti, \Tipos)$ by \cref{sec:contrastive-fw}\;
            $\batchP \gets \batchP \cup \{\Ti, \Tipos\}$\;
        }

        \ForEach{$\Ti \in \batchP$}{
            $\hi \gets \TEnc(\Ti)$\tcp*[l]{Table embedding (\cref{sec:contrastive-att})}
            \tcp{Two-pronged negative sampling}
        Get $\Nhardi$ of $\Ti$ by \cref{eq:latent-set,eq:hard-neg}\;
        }
 
        Get batch loss $\mathcal{L}$ by \cref{eq:single-loss,eq:full-loss}
        
        $\TEnc \gets \text{back-propagate}(\TEnc, \mathcal{L})$\;
        }
}
\KwRet{$\TEnc$}
\end{algorithm}

\subsection{Training Procedure}\label{sec:training-indexing}

\cref{alg:contrastive-learning} outlines the training procedure. Given a data lake $\lake$ of tables, each epoch begins by randomly partitioning $\lake$ into batches (Lines 1-2). 
For each batch $\batch$, we construct its augmented batch $\batchP$ by generating a positive table pair $(\Ti, \Tipos)$ for each table $\Ti$ in $\batch$ using the positive table pair construction described in \cref{sec:contrastive-fw}, and include both $\Ti$ and $\Tipos$ in $\batchP$ (Lines 3-7). 
Then, for each table $\Ti$ in $\batchP$, we obtain its table embedding $\hi$ using the attentive table encoder $\TEnc$ described in~\cref{sec:contrastive-att} (Lines 8–9).
Next, we perform two-pronged negative table sampling  in \cref{eq:latent-set,eq:hard-neg} to obtain the hard negative set $\Nhardi$ for each $\Ti \in \batchP$ (Line 10).

Notably, the two-pronged negative table sampling (Line 10) is performed within each batch $\batchP$ in the for loop at Line 8, using only the current table embeddings computed in this batch (Line 9). This ensures that only information available within the current batch is used, thereby avoiding any information leakage or circular dependencies.
The loss $\mathcal{L}$ is then computed at Line 11, and the model is updated via back-propagation with the Adam optimizer (Line 12).
Finally, the trained table encoder $\TEnc$ is returned (Line 13).

After training, we use the trained table encoder $\TEnc$ to compute the table embedding $\hi = \TEnc(\Ti)$ for each table $\Ti$ in the data lake $\lake$, resulting in the set of table embeddings $\batchEmb = \{\hi \mid \Ti \in \lake\}$.  
Each table embedding is a multi-dimensional vector, and the similarity between two table embeddings quantifies the unionability of their corresponding tables. Notably, the number of table embeddings equals the number of tables in $\lake$, whereas existing methods generate column embeddings for all columns in the lake.

For efficient approximate nearest neighbor search over embeddings, various vector indexing methods are available~\cite{LiZSWLZL20,PanWL24,WangXY021}. Prior work~\cite{starmie,liftus} uses HNSW~\cite{hnsw} for column embeddings. 
The choice of index is orthogonal to table union search. For fair comparison, we use the same index, but index our table embeddings.

\section{\our Online Search} \label{sec:online}

Given a large data lake $\lake$, searching for unionable tables with a query table $\Tq$ is computationally expensive if the entire lake is exhaustively inspected to obtain the top-$k$ results.

A common approach is the filter-and-refine paradigm. 
Existing methods~\cite{liftus,starmie} follow a column-centric strategy: they filter candidate tables by column-level unionability scores to any column in $\Tq$, then refine by aggregating scores of highly-matched columns for top-$k$ selection.
In the filtering stage, since a query table $\Tq$ often has multiple columns, and each column may match columns from many tables, prior work retrieves a large candidate set.
During refinement, they aggregate column unionability scores of highly-matched columns, overlooking the holistic table-level semantics across all columns.
 
We argue that both the semantics of entire tables and   individual columns are essential for accurate table union search.
Therefore, we design our search process in a table-centric manner: only when a table is sufficiently promising at the table level is it worthwhile to examine its columns in detail with the query $\Tq$. \cref{alg:cand-retrival} outlines the online search process of our method \our. 

First, we propose a table-centric adaptive candidate retrieval method (\cref{sec:retrieval}) that leverages the table embeddings from \cref{sec:training-indexing} to efficiently filter out irrelevant tables and obtain a compact, controllable candidate pool $\candset$ of likely unionable tables based on their table-level unionability scores $\scOne(\Tq, \Ti)$ to the query $\Tq$. Since this process operates on table embeddings—one per table—it enables flexible adjustment of the candidate pool size according to the distribution of $\scOne(\Tq, \Ti)$ scores, rather than being constrained by the number of columns in $\Tq$ as in existing methods. Our approach is distribution-aware: it adaptively determines the candidate pool size based on the observed unionability score distribution for the query, resulting in a concise yet high-coverage candidate set.

Then, over the compact candidate pool $\candset$, we perform a final dual-evidence reranking step (\cref{sec:rerank}) that combines the table-level unionability score $\scOne(\Tq, \Ti)$ from table embeddings with a lightweight column alignment score $\scTwo(\Tq, \Ti)$ to compute the final unionability score $\score(\Tq, \Ti)$ for each candidate. This approach effectively integrates table-level  and column-level semantics to refine the top-$k$ results.

\begin{algorithm}[t] 
\caption{\our Search Process}
\label{alg:cand-retrival}
\KwIn{Query table $\Tq$,    table embeddings $\batchEmb$,   absolute threshold $\tau_{\text{abs}}$, drop threshold $\tau_{\text{drop}}$, result size $k$}
\KwOut{Top-$k$ result $\lake_q$}
\tcp{Table-Centric Adaptive Candidate Retrieval. (\cref{sec:retrieval})}
$\hq\gets \TEnc(\Tq)$\;

$\Cinit \gets$ Get the top-$3k$ tables for $\Tq$ by nearest neighbor search on table embeddings $\batchEmb = \{\hi \mid \Ti \in \lake\}$, ranked in descending order of $\scOne(\Tq, T_i)$\;

$i_{\text{cut}} \gets k$\;
\For{$i \gets k+1$ \textbf{ to } $3k$}{
    \If{$\scOne(\Tq, T_i) < \tau_{\text{abs}}$ \textbf{or} 
        $\scOne(\Tq, T_{i-1}) - \scOne(\Tq, T_i) > \tau_{\text{drop}}$}{
        $i_{\text{cut}} \gets i-1$\;
        \textbf{break}\;
    }
}

$\candset \gets \Cinit[:i_{\text{cut}}]$\;

\tcp{Dual-Evidence Reranking (\cref{sec:rerank})}
\ForEach{candidate table $\Ti \in \candset$}{
    Compute $\scTwo(\Tq, \Ti)$ by \cref{eq:column-simi}\;
    $\score(\Tq, \Ti) \gets \scOne(\Tq, \Ti) + \scTwo(\Tq, \Ti)$\;
}
Sort $\candset$ by $\score(\Tq, \Ti)$ in descending order\;
$\lake_q \gets \candset[:k]$\;
\Return{$\lake_q$}\;
\end{algorithm}

\subsection{Table-Centric Adaptive Candidate Retrieval}
\label{sec:retrieval} 

Given the table embeddings $\batchEmb = \{\hi \mid \Ti \in \lake\}$ from \cref{sec:contrastive}, we can efficiently compute the table-level unionability score $\scOne(\Tq, \Ti)$ for each $\Ti$ and retrieve the top-ranked candidates.

However, a fixed-size candidate pool is suboptimal: if too large, it includes many irrelevant tables; if too small, it may miss unionable ones. 
Moreover, the effective pool size varies by query, as the number of promising unionable tables differs across queries.
Thus, candidate retrieval should adaptively adjust the pool size based on the unionability score distribution for each query.

Hence, we propose a table-centric adaptive candidate retrieval method that constructs a high-quality candidate pool $\candset$ for a query table $\Tq$, with the pool size adaptively determined by the distribution of table-level unionability scores $\scOne$.

In \cref{alg:cand-retrival}, Lines 1–8 detail the construction of $\candset$.
Given a query table $\Tq$, we first compute its table embedding $\hq$ (Line 1), and then perform a single nearest-neighbor search over the indexed table embeddings $\batchEmb$ to retrieve an initial set $\Cinit$ containing the top-$3k$ tables with the highest table-level unionability scores $\scOne(\Tq, \Ti)$ (Line 2).
$\Cinit$ serves as the initial pool from which we adaptively determine the candidate pool $\candset$.
Empirically, we set $|\Cinit| = 3k$, which is sufficiently large to cover potentially unionable tables.
This is because our table embeddings, as designed in \cref{sec:contrastive}, effectively encode union-oriented semantics, so the identified top-$3k$ tables are very likely to cover all truly unionable tables, while those beyond this range are typically irrelevant and not worth further consideration. Experiments in~\cref{fig:vary3k} show that varying $|\Cinit|$ beyond $3k$ yields negligible gains. Moreover, this step is efficient and accounts for only a negligible fraction of query time.

Lines 3-8 adaptively determine the candidate pool $\candset$ by scanning the ranked list $\Cinit$ and selecting a cutoff position $i_{\text{cut}}$ based on the distribution of table-level unionability scores $\scOne(\Tq, \Ti)$. This cutoff separates likely unionable tables from less relevant ones.
We set the initial cutoff $i_{\text{cut}}$ to $k$ to ensure at least $k$ candidates. Intuitively, if $\scOne(\Tq, \Ti)$ drops sharply or falls below a threshold, this marks a natural boundary in the score distribution, beyond which tables are unlikely to be unionable. 
To determine the cutoff position, we scan the tables in $\Cinit$ in descending order of $\scOne(\Tq, \Ti)$. The scan stops when either of the following conditions is met: (1) the score $\scOne(\Tq, \Ti)$ of the current table drops below an absolute threshold $\tau_{\text{abs}}$, or (2) the difference in scores between two consecutive tables exceeds a threshold $\tau_{\text{drop}}$. We then set the cutoff position $i_{\text{cut}}$ to the previous table's position (Lines 4–7). The threshold $\tau_{\text{abs}}$ specifies the minimum unionability score required for a table to be considered as a candidate, while $\tau_{\text{drop}}$ identifies a significant drop in the score distribution, which separates unionable tables from less relevant ones.
In \cref{sec:exp}, we use the same threshold settings for all datasets in our experiments and also vary them  to assess their impact and provide practical guidance.
Then, the candidate pool $\candset$ consists of the top tables in $\Cinit$ up to the cutoff $i_{\text{cut}}$ (Line 8).

\header
{\bf Discussion.} (i) Our candidate retrieval operates solely on table embeddings—one vector per table—so the candidate pool size is directly controllable and does not depend on the number of columns in the query table, in contrast to existing column-centric methods. (ii) It is possible to directly return the top-$k$ tables ranked by $\scOne(\Tq, \Ti)$ (i.e., by retrieving the top-$k$ in Line 2 of \cref{alg:cand-retrival}) as the final result, without further refinement. As shown in our ablation study in~\cref{tab:ablation}, this approach already outperforms existing methods, highlighting the effectiveness of the table embeddings developed in \cref{sec:contrastive}.

\subsection{Dual-Evidence Reranking}
\label{sec:rerank}

Nevertheless, it is well established that  column-level semantics are also crucial for accurate table union search. Thus, we incorporate column-level information only in this final step, using a dual-evidence reranking approach that combines table-level unionability and column alignment over the candidate pool $\candset$.

Specifically, the dual-evidence reranking computes the final table unionability score $\score(\Tq, \Ti)$ for each candidate table $\Ti \in \candset$ by integrating two complementary signals: (i) the table-level unionability score $\scOne(\Tq, \Ti)$ derived from table embeddings, and (ii) a column alignment score $\scTwo(\Tq, \Ti)$ that captures fine-grained column compatibility between $\Tq$ and $\Ti$, as described below.
For each table $T$, we precompute and store the FastText embeddings~\cite{fasttext} of its columns $\cj$ during the offline stage. FastText generates word embeddings by representing each word as a bag of character n-grams and aggregating the corresponding n-gram embeddings, efficiently capturing subword information. 

FastText can be used off-the-shelf without additional  training and is computationally efficient and providing effective performance for our purposes. While other models such as BERT could also be employed, as shown in \cref{tab:fasttext-bert-rerank} of experiments, FastText achieves strong performance with greater efficiency in our method.

For each unique value $v$ in the value set $\Vj$ of column $\cj$, we obtain its FastText embedding $\FTenc(v)$. We then compute the FastText-based column embedding $\ejemb$ for $\cj$ by aggregating these value embeddings, weighted by the frequency $f(v)$ of $v$ in the column:
\begin{equation}\label{eq:fasttext-embed}
\ejemb = \frac{\sum_{v \in \Vj} f(v) \cdot \FTenc(v)}{\sum_{v \in \Vj} f(v)}
\end{equation}

During online search, we first compute the FastText-based column embedding $\eremb$ for each column $\crq$ of the query table $\Tq$ using \cref{eq:fasttext-embed}. For each candidate table $\Ti$ in the pool $\candset$, we compute the column alignment score $\scTwo(\Tq, \Ti)$ as defined in \cref{eq:column-simi}, by aligning each query column $\crq$ to its most similar column in $\Ti$:
\begin{equation}\label{eq:column-simi}
\scTwo(\Tq, \Ti)= \frac{1}{|\Tq|} \sum_{\crq \in \Tq} \max_{\cj \in \Ti} \phi(\eremb, \ejemb),
\end{equation}
where $\phi(\cdot, \cdot)$ denotes cosine similarity, and $|\Tq|$ is the number of columns in $\Tq$.

Our final table unionability score $\score(\Tq, \Ti)$ is defined as the sum of the table-level unionability score $\scOne(\Tq, \Ti)$ and the column alignment score $\scTwo(\Tq, \Ti)$, as shown in \cref{eq:final-score}. 

Note that both $\scOne(\Tq, \Ti)$ and $\scTwo(\Tq, \Ti)$ are in the range $[0,1]$ and exhibit similar means and standard deviations within each dataset in the experiments. For example, on the Wiki Union dataset, the mean (std) of $\scOne$ is $0.780 (0.125)$, while that of $\scTwo$ is $0.795 (0.155)$. Therefore, we simply sum them to obtain the final score. This parameter-free approach already achieves strong performance in our experiments. More sophisticated combinations, such as a weighted sum with a tunable hyperparameter, could be explored in the future.
\begin{equation}\label{eq:final-score}
\score(\Tq, \Ti) = \scOne(\Tq, \Ti) + \scTwo(\Tq, \Ti).
\end{equation}

As shown in Lines 9–14 of \cref{alg:cand-retrival}, for each candidate table $\Ti$ in $\candset$, we compute its column alignment score $\scTwo(\Tq, \Ti)$ using \cref{eq:column-simi}, and then combine it with $\scOne(\Tq, \Ti)$ to obtain the final table unionability score $\score(\Tq, \Ti)$. The candidates are then reranked by this score at Line 12, and the top-$k$ tables are returned as the final result $\lake_q$ at Line 14. This dual-evidence reranking incorporates the column alignment score $\scTwo(\Tq, \Ti)$ as a complementary signal to the table-level unionability score $\scOne(\Tq, \Ti)$, and is applied only to the candidate tables in $\candset$, thereby further improving result quality.

\section{Experiments} 
\label{sec:exp}
We conduct extensive experiments using widely-adopted benchmark datasets to evaluate the effectiveness and efficiency of our proposed method \our with state-of-the-art baselines. 

\subsection{Experiment Setup}
\label{sec:exp-setup}

\begin{table}[!t]
  \centering 
  \caption{Dataset Statistics}  
  \label{tab:dataset}
  \vspace{-3mm}
  \resizebox{0.96\linewidth}{!}{
  \setlength{\tabcolsep}{5pt}
    \begin{tabular}{lcccc}
      \toprule
      Benchmark & \# Tables & \# Cols & Avg \# Cols & Avg \# Rows \\
      \midrule
      \santosSmall  & 550      & 6,322      & 11.49 & 6,921    \\
      \TUSSmall     & 1,530    & 14,810     & 9.68  & 4,466    \\
      \TUSLarge     & 5,043    & 54,923     & 10.89 & 1,915    \\
      \wiki    & 40,752   & 106,744    & 2.62  & 51       \\
      \hline
      \santosLarge  & 11,090   & 123,477    & 11.13 & 7,675    \\
\WDC & 1,000,000       & 5,438,291  & 5.44  & 14.27    \\
      \hline
    \end{tabular}
    \vspace{-1mm}
  }
\end{table}

\begin{table*}[!t]
  \centering
  \caption{MAP@k, P@k, and R@k results on all benchmarks with ground truth, where k=10 for \santosSmall, k=60 for the TUS benchmarks and k=40 for \wiki. 
  {\em Avg. Rank} in the last column represents, for each method, the average of its ranks among all methods across all metrics and datasets.
  The best is in bold, and the second best is underlined.} 
  \label{tab:overall}
  \vspace{-3.3mm}
  \resizebox{0.98\textwidth}{!}{
    \begin{tabular}{lccccccccccccc}
    \toprule
    \multirow{2}[4]{*}{Method} & \multicolumn{3}{c}{\santosSmall} & \multicolumn{3}{c}{\TUSSmall} & \multicolumn{3}{c}{\TUSLarge} & \multicolumn{3}{c}{\wiki} & \multirow{2}[4]{*}{Avg. Rank} \\
    \cmidrule{2-13}    & $MAP@k$ & $P@k$ & $R@k$ & $MAP@k$ & $P@k$ & $R@k$ & $MAP@k$ & $P@k$ & $R@k$ & $MAP@k$ & $P@k$ & $R@k$ & \\
    \hline
    \dthreeL   & 0.5595 & 0.5100 & 0.4167 & 0.7916 & 0.7762 & 0.2147 & 0.5155 & 0.5102 & 0.1430 & 0.0905 & 0.0575 & 0.0684 & 6.42\\
    \sherlock & 0.8082 & 0.6860 & 0.5080 & 0.9592 & 0.9220 & 0.2886 & 0.7600 & 0.5897 & 0.1296 & 0.2960 & 0.2418 & 0.2017 & 5.25 \\
    \sato     & 0.8822 & 0.8280 & 0.6077 & 0.9462 & 0.9302 & 0.2928 & 0.8924 & 0.8228 & 0.1924 & \underline{0.5263} & \underline{0.4495} & \underline{0.3701} & 3.83 \\
    \santos   & 0.9446 & 0.9160 & 0.6820 & 0.8892 & 0.8406 & 0.2708 & - & - & - & - & - & -  & 5 \\
    \starmie  & 0.9644 & 0.9360 & 0.6926 & 0.9672 & 0.9342 & 0.3100 & 0.9113 & 0.8508 & 0.2199 & 0.4485 & 0.3930 & 0.3115 & 3 \\
    \liftus  & \underline{0.9730} & \underline{0.9600} & \underline{0.7158} & \underline{0.9709} & \underline{0.9381} & \underline{0.3112} & \underline{0.9662} & \underline{0.9357} & \underline{0.2397} & 0.3438 & 0.2850 & 0.2331 & 2.5 \\
    \midrule
   \our   & \textbf{0.9860} & \textbf{0.9720} & \textbf{0.7290} & \textbf{0.9936} & \textbf{0.9876} & \textbf{0.3299} & \textbf{0.9844} & \textbf{0.9728} & \textbf{0.2516} & \textbf{0.6537} & \textbf{0.6238} & \textbf{0.5066} & \textbf{1}\\
    \hline
    \end{tabular} 
    \vspace{-1.5mm} 
    }
\end{table*} 

{\bf Datasets.} We use six benchmark datasets adopted in~\cite{starmie,liftus,santos}, with their statistics summarized in \cref{tab:dataset}.  
The first four datasets provide ground truth for unionable tables and are used to evaluate both effectiveness and efficiency, while the last two datasets lack ground truth and are used only for online efficiency evaluation, following~\cite{starmie,liftus}. 
The \santosSmall benchmark~\cite{santos} comprises 550 real tables from open datasets in Canada, the UK, the US, and Australia, with 50 query tables. 
The \TUSSmall and \TUSLarge benchmarks~\cite{nargesian2018table} contain 1,530 and 5,043 tables, respectively, generated from Canadian open data. We randomly select 150 and 100 query tables for \TUSSmall and \TUSLarge, respectively~\cite{santos,starmie,liftus}.
The \wiki benchmark~\cite{srinivas2023lakebench} is constructed from Wikidata~\cite{vrandevcic2014wikidata} and contains 40,752 tables. 
Originally designed for supervised table unionability prediction~\cite{srinivas2023lakebench}, it provides ground-truth labels for unionable table pairs. 
To adapt this dataset for unsupervised table union search, we identify all unionable tables for each table based on ground truth and randomly select 100 tables, each with more than 40 unionable tables, as query tables. 
Following prior work~\cite{starmie,liftus}, the \santosLarge and \WDC~\cite{lehmberg2016large} datasets are used only for online efficiency evaluation, as they lack ground truth. \santosLarge includes 80 randomly selected query tables~\cite{santos,starmie,liftus}, while \WDC consists of one million tables randomly sampled from the WDC web tables corpus, with 50 randomly selected query tables.
These six datasets vary in size, column and row counts, and table widths, providing a comprehensive evaluation of different methods.

\header 
{\bf Baselines.} We compare \our with state-of-the-art methods, including \starmie~\cite{starmie}, \liftus~\cite{liftus}, \santos~\cite{santos}, \dthreeL~\cite{d3l}, \sherlock~\cite{sherlock}, and \sato~\cite{sato}. \dthreeL was originally designed  using features such as table content and column names; following~\cite{santos,starmie,liftus}, we configure it without access to column names for fair comparison. \sherlock and \sato produce column representations; to enable table union search, we follow the evaluation protocol in~\cite{starmie}, aligning their online query processing with that of \starmie since all three methods produce column embeddings.

\header 
\textbf{Implementations.} 
For all competitors, we use their publicly available codebases. 
For fair comparison, we apply HNSW~\cite{hnsw} to index the column embeddings produced by \starmie, \sato, \liftus, and \sherlock, and the table embeddings produced by our \our. Baselines are configured with the recommended parameter settings from their respective papers.
We implement \our in Python using PyTorch and the Transformers library~\cite{wolf2020transformers}, adopting BERT~\cite{bert} as the pretrained language model.

We evaluate with maximum input sequence lengths set to be 256 and 512 tokens for all language models, and unless otherwise specified, we use 256 tokens by default, following prior work~\cite{starmie}. Our method \our uses the same parameter settings across all datasets, including $\gamma=0.9$, $\tau_{\text{abs}}=0.5$, and $\tau_{\text{drop}}=0.2$, demonstrating robust performance across datasets; we also analyze the impact of varying parameters. For each dataset, query tables are {\em excluded} from all offline processing, including training, to ensure no data leakage. Since no ground-truth labels are used during training, data splitting is unnecessary. We use a batch size of 64. Experiments are conducted on a server with an Intel Xeon Platinum 8370C CPU at 2.80GHz and an NVIDIA RTX Ada 6000 GPU.

\header 
\textbf{Evaluation Metrics.}
For effectiveness~\cite{santos,starmie,liftus}, we use Mean Average Precision at $k$ ($MAP@k$), Precision at $k$ ($P@k$), and Recall at $k$ ($R@k$) to evaluate. 
Given a query table $\Tq$, let $\lake_{gt}$ be its truly unionable tables and $\lake_{q}$ the set of returned top-$k$ tables by a method. The metrics are defined as:
$
P@k = \frac{|\lake_{q} \cap \lake_{gt}|}{k}, R@k = \frac{|\lake_{q} \cap \lake_{gt}|}{|\lake_{gt}|},  MAP@k = \frac{1}{k} \sum_{i=1}^{k} P@i$.
$P@k$ measures the proportion of top-$k$ returned tables that are truly unionable with $\Tq$; $R@k$ measures the proportion of all truly unionable tables included in the top-$k$; and $MAP@k$ averages $P@i$ over $i=1$ to $k$. 
We also report the average rank (Avg. Rank) of each method, computed as its average ranking across all metrics and datasets.
For efficiency, we report offline processing time, online processing time, and embedding index size. 
Following~\cite{santos,starmie,liftus}, we set $k=60$ for TUS datasets, $k=10$ for \santosSmall, \santosLarge, and \WDC, and $k=40$ for \wiki.

\begin{figure*}[t]
    \centering
    \includegraphics[width=0.78\textwidth]{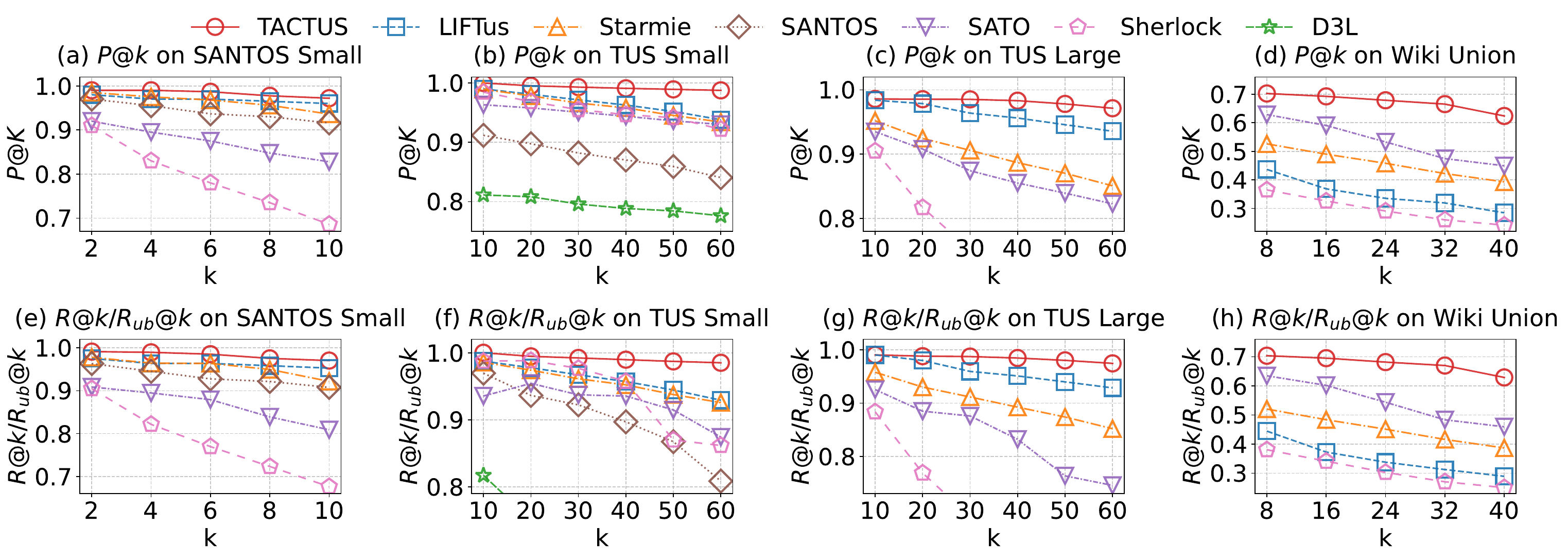}

        \caption{Precision $P@k$ and Relative Recall $R@k/R_{ub}@k$  with Varied $k$}
        \label{fig:overall_figure}
        \vspace{-2mm}
\end{figure*}

\subsection{Effectiveness Evaluation}
\label{sec:effectiveness}
\cref{tab:overall} reports the effectiveness results of \our and the baselines on the four datasets with ground truth.  
\santos is not reported on \TUSLarge and \wiki due to its reliance on additional knowledge, which is unavailable for these datasets~\cite{santos,starmie}.

Our method \our consistently achieves the best performance across all datasets and metrics, with an average rank of 1.0. 
For example, on \TUSSmall, \our attains $MAP@k$ of 99.36\%, $P@k$ of 98.76\%, and $R@k$ of 32.99\%, outperforming the second-best method \liftus by 2.27\%, 4.95\%, and 1.87\%, respectively, and \starmie by 2.64\%, 5.34\%, and 1.99\%, respectively.
Similar trends hold for larger datasets, where \our achieves the best results. 
For example, on \TUSLarge, \our achieves $P@k$ of 97.28\%, exceeding \liftus by 3.71\% and \starmie by 12.20\%. 
These results confirm the effectiveness of our table-centric embedding techniques (\cref{sec:contrastive}) and search process (\cref{sec:online}). 
The table embeddings learned by \our capture table-level unionability semantics, enabling the table-centric adaptive candidate retrieval in \cref{sec:retrieval} to obtain high-quality candidates that are likely to be unionable with the query table. 
The dual-evidence reranking in \cref{sec:rerank} further refines these candidates by jointly leveraging table- and column-level evidence, leading to superior performance.

On the \wiki dataset, all methods exhibit lower performance than on the other datasets, as tables in \wiki typically contain fewer columns and rows, providing limited contextual information for table union search. Nevertheless, \our significantly outperforms all baselines, achieving $MAP@k$ of 65.37\%, $P@k$ of 62.38\%, and $R@k$ of 50.66\%, which are 12.74\%, 17.43\%, and 13.65\% higher than the runner-up \sato, respectively. Column-centric methods such as \liftus and \starmie are less effective on \wiki because the limited context constrains the quality of the column embeddings and candidate retrieval. In contrast, \our leverages holistic table-level semantics to produce table embeddings as described in \cref{sec:contrastive}, which is more effective for large, sparse benchmarks like \wiki. Moreover, our table-centric adaptive candidate retrieval and dual-evidence reranking techniques in \cref{sec:online} further enhance online search effectiveness, leading to substantial improvements over the baselines on \wiki.

We further vary $k$ to evaluate precision and recall. 
Figures~\ref{fig:overall_figure}(a,b,c,d) report precision $P@k$ as $k$ increases.
Precision decreases for all methods as $k$ grows; \our consistently maintains a clear advantage across all settings, especially at larger $k$. 
For example, in \cref{fig:overall_figure}(b) (\TUSSmall), at $k=40$, \our achieves 99.10\% precision, which is 2.82\% higher than \liftus; on \wiki in \cref{fig:overall_figure}(d), at $k=32$, \our attains 66.53\% precision, 19.07\% higher than \sato (47.46\%).

For recall, note that when $k$ is smaller than the number of ground-truth unionable tables $|\lake_{gt}|$ for a query $\Tq$, recall cannot reach 100\%. 
The upper bound, i.e., the best possible recall, is $R_{ub}@k = \frac{\min(k, |\lake_{gt}|)}{|\lake_{gt}|}$. 
For example, on \TUSLarge with $k=60$, $R_{ub}@k$ is as low as 0.2581; for even smaller $k$ (e.g., $k=10$), $R_{ub}@k$ becomes much lower. 
This makes the $R@k$ curves of different methods difficult to distinguish, as also observed in~\cite{starmie,liftus}.
To better visualize recall trends, we therefore adopt a relative recall metric, $R@k/R_{ub}@k$, which normalizes the achieved recall by the upper bound $R_{ub}@k$. As shown in Figures~\ref{fig:overall_figure}(e,f,g,h), relative recall decreases for all methods as $k$ increases, which matches the intuition that a larger $k$ makes it harder to retrieve all unionable tables. 
Across all $k$, \our achieves the highest relative recall, with the gap widening as $k$ grows. 
These results further validate the effectiveness of our table embedding techniques in \cref{sec:contrastive} and the online search techniques in \cref{sec:online}.

\subsection{Efficiency Evaluation}
\label{sec:efficiency}

\begin{figure}
    \centering
    \includegraphics[width=0.86\linewidth]{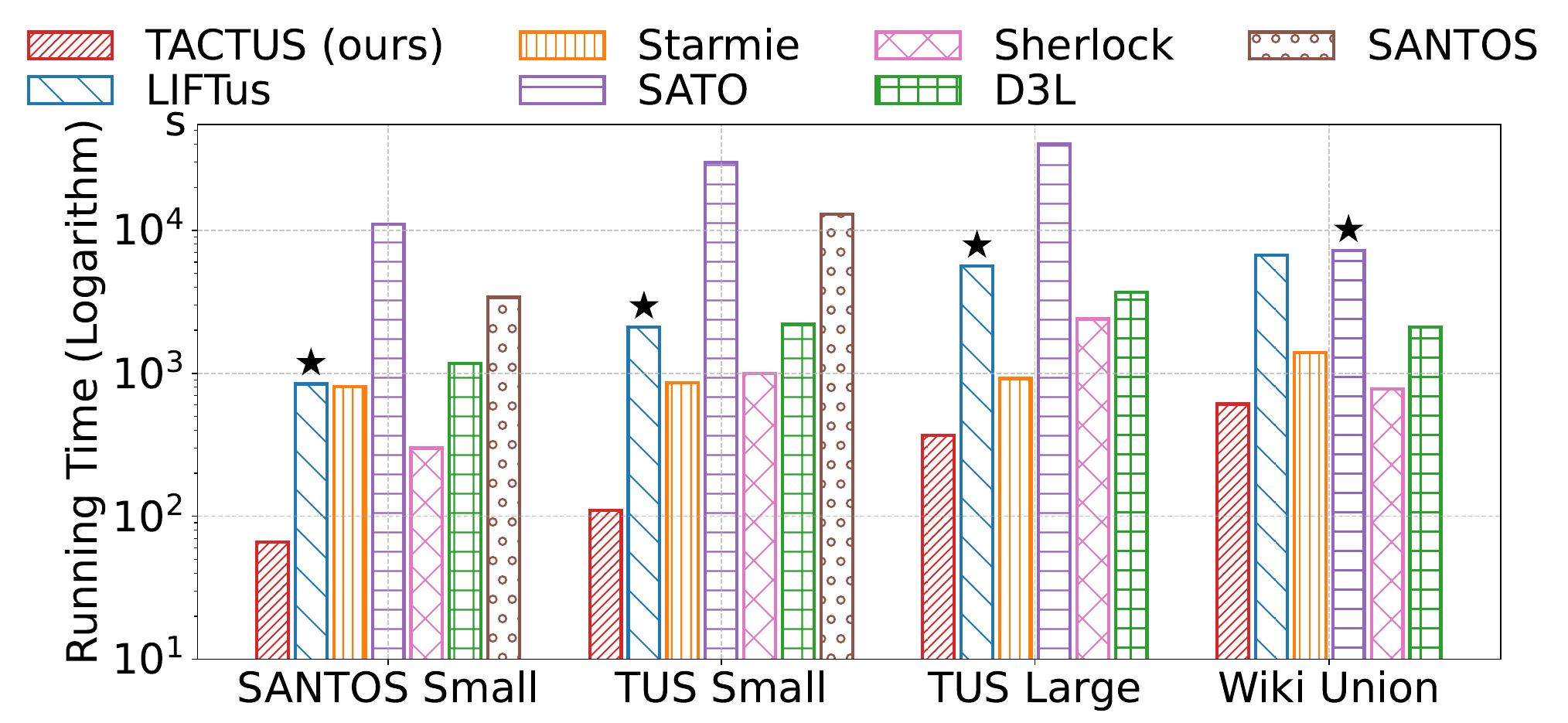}
    \caption{Offline Processing Time in Seconds (The baseline with runner-up effectiveness in \cref{tab:overall} is marked with $\star$)}
    \label{fig:offline-effi}
        \vspace{-2mm}
\end{figure}

\begin{figure}
    \centering
    \includegraphics[width=0.95\linewidth]{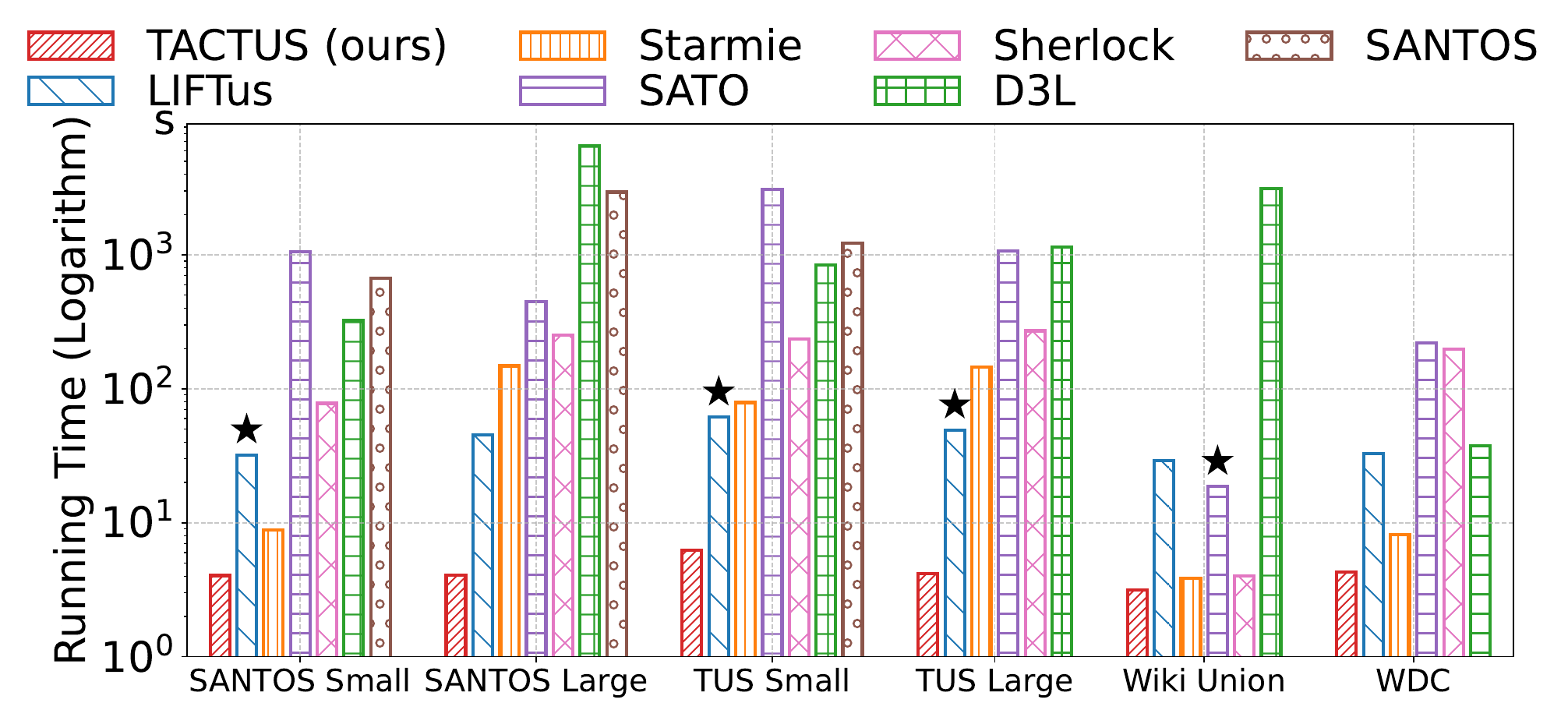}
    \caption{Online Processing Time in Seconds (The baseline with runner-up effectiveness in \cref{tab:overall} is marked with $\star$)}
    \label{fig:al-online-effi}
        \vspace{-2mm} 

\end{figure}

The offline time includes data preprocessing, model training, embedding generation, and index construction. We report the offline processing time of \our and the baselines on the four datasets with ground truth used in the effectiveness evaluation. Note that \santosLarge and \WDC do not require offline training and are only used for online efficiency evaluation.

As shown in \cref{fig:offline-effi} (log-scale), \our is significantly faster than all baselines, including those with runner-up effectiveness in \cref{tab:overall}, often by more than an order of magnitude. 
For example, on \santosSmall, \our completes offline processing in only {66} seconds, which is $12.8\times$ faster than \liftus ({842} seconds) and $12.3\times$ faster than \starmie ({811} seconds), the two strong baselines in \cref{tab:overall}. On \wiki, \our finishes offline processing in {610} seconds, which is $11.8\times$ faster than \sato ({7217} seconds), the baseline with the second-best effectiveness in \cref{tab:overall}; while \starmie is relatively efficient on \wiki, its effectiveness in \cref{tab:overall} is lower.
In addition, \starmie is more efficient than \liftus due to the latter's additional multi-aspect computation per column. Among the baselines, \sherlock is relatively fast, but as shown in \cref{tab:overall}, its effectiveness is lower, with an average rank of 5.25.
The offline efficiency of \our primarily stems from its table-centric offline processing, as described in \cref{sec:contrastive}, which contrasts with existing methods that require building and processing column embeddings. 

The online processing time for a method mainly includes two steps:  encode query tables and columns into embeddings; search for top-$k$ unionable tables. The first step typically dominates the total time, though prior work often reports only the second step.

In \cref{fig:al-online-effi} (log-scale), we report the total online processing time for all methods across all datasets. 
\our consistently achieves faster online processing than all baselines, often by more than an order of magnitude. 
For example, on \TUSLarge, \our processes all queries in just {$4.18$} seconds, which is {$11.80\times$} faster than \liftus ({49.33} seconds) and {$34.95\times$} faster than \starmie ({146.11} seconds), and significantly faster than the other baselines. 
On both \wiki and \WDC, which have fewer columns per table, \our still maintains a clear efficiency advantage. On \wiki, \sato, the runner-up in \cref{tab:overall}, is {6$\times$} slower than \our; while \starmie and \liftus are comparable in efficiency, they yield lower effectiveness as reported in \cref{tab:overall}.
The online efficiency of \our demonstrates that our table-centric adaptive candidate retrieval in \cref{sec:retrieval} can quickly retrieve a small number of high-quality candidates for reranking to obtain the final top-$k$ results, as described in \cref{sec:online}. The experiments in \cref{tab:candidate} in \cref{sec:expAnalysis} further shows that \our retrieves fewer candidates than the baselines while achieving higher effectiveness, validating the strength of our table-centric modeling and search techniques in \cref{sec:contrastive,sec:online}.

\subsection{Experimental Analysis} \label{sec:expAnalysis}

\noindent 
\textbf{Comparison with Baselines using Aggregated Column Embeddings as Table Embeddings.}
For a table $\Ti$, existing methods~\cite{starmie,liftus} produce column embeddings. A straightforward approach to obtain a table embedding is to aggregate its column embeddings, e.g., via average pooling. To further demonstrate the effectiveness and necessity of our table embedding techniques in \cref{sec:contrastive}, we adapt \starmie and \liftus as follows: aggregate their column embeddings to form table embeddings for candidate retrieval, and rerank the retrieved candidates using their original unionability scoring functions. We refer to these variants as \starmieAgg and \liftusAgg, respectively.
As shown in \cref{tab:aggregating-column}, \our consistently outperforms both \starmieAgg and \liftusAgg across all datasets. This is because the baseline column embeddings are optimized for column-level unionability, and simple aggregation cannot capture the holistic table semantics required for table-level unionability. These results highlight the effectiveness of our table-centric embedding techniques in \cref{sec:contrastive}.

\header 
\textbf{Performance with 512 Tokens.} \cref{tab:bert-512} reports the performance when setting the maximum input sequence length of BERT to 512 tokens. We observe that (i) all methods achieve similar performance to that in \cref{tab:overall} with 256 tokens, with only minor differences; and (ii) \our consistently maintains its performance advantage over the baselines. These results demonstrate the robustness of our method and justify the use of 256 tokens as the default setting, following common practice~\cite{starmie,liftus}.

\header 
\textbf{Ablation Study.} We evaluate the contribution of each major component in \our by considering the following ablated variants: (i) disabling the dual-evidence reranking in Section~\ref{sec:rerank} and using only $\scOne(\Tq,\Ti)$ as the final unionability score to select the top-$k$ tables ({w/o rerank}); (ii) removing the two-pronged negative sampling in Section~\ref{sec:contrastive-fw} ({w/o two-pronged}); (iii) replacing the adaptive candidate retrieval in Section~\ref{sec:retrieval} with a fixed candidate pool size of $3k$ ({w/o adaptive}); and (iv) substituting the attentive aggregation in Section~\ref{sec:contrastive-att} with simple average pooling ({w. avgPooling}).
As shown in \cref{tab:ablation}, each component contributes to the overall performance of \our across all datasets. For example, the reranking module improves MAP@60 on \TUSSmall from 97.95\% to 99.36\%, a 1.41\% absolute gain.
Notably, \our (\textit{w/o rerank}), i.e., using only the table-centric candidate retrieval in \cref{sec:retrieval} to get the top-$k$ tables, already outperforms all baselines in \cref{tab:overall} in MAP@k on all datasets, validating the effectiveness of our table-centric designs in \cref{sec:contrastive,sec:retrieval}. 
Moreover, two-pronged negative sampling is particularly effective for \santosSmall and \TUSSmall, while adaptive candidate retrieval and attentive aggregation are more beneficial on \wiki and \TUSLarge.

\begin{table}[!t]
  \caption{Performance of baseline variants using aggregated column embeddings as table embeddings}
  \label{tab:aggregating-column}
  \vspace{-3.5mm}
  \resizebox{0.92\linewidth}{!}{
              \setlength{\tabcolsep}{5pt}
    \begin{tabular}{lcccc}
    \toprule
    {{MAP@k}} & \santosSmall & \TUSSmall & \TUSLarge & \wiki \\
    \hline
    {\our}     & \textbf{0.9860}    & \textbf{0.9936}    & \textbf{0.9844}    & \textbf{0.6537} \\
    {\starmieAgg}     & 0.9768             & 0.9511             & 0.8981             &  0.4380 \\
    {\liftusAgg}      & 0.9452             & 0.9227             & 0.9080             &  0.3277 \\
    \hline
    \end{tabular}
  }
  \vspace{-2.5mm}
\end{table}

\begin{table}[!t]
  \centering
  \caption{MAP@k with maximum token length 512}
  \label{tab:bert-512}
  \vspace{-3.5mm}
  \resizebox{0.92\linewidth}{!}{
              \setlength{\tabcolsep}{5pt}
    \begin{tabular}{lcccc}
    \toprule
    {{Method}} & \santosSmall & \TUSSmall & \TUSLarge & \wiki \\
    \hline
    {\our}     & \textbf{0.9895} & \textbf{0.9951} & \textbf{0.9830} & \textbf{0.6514} \\ 
    {\starmie} & 0.9618 & 0.9670 & 0.9143 & 0.4505 \\ 
    {\liftus}  & 0.9736 & 0.9718 & 0.9656 & 0.3438 \\ 
    \hline
    \end{tabular}
  }
  \vspace{-2.5mm}
\end{table}

\begin{table}[t]
  \centering
  \caption{Ablation Study (MAP@k)}
  \label{tab:ablation}
  \vspace{-3.5mm}
  \resizebox{\linewidth}{!}{
              \setlength{\tabcolsep}{2pt}
    \begin{tabular}{lcccc}
    \toprule
    {{Variant}} & \santosSmall & \TUSSmall & \TUSLarge & \wiki \\
    \hline
    {\our} & \textbf{0.9860} & \textbf{0.9936} & \textbf{0.9844} & \textbf{0.6537} \\
    {\our w/o rerank} & 0.9788 & 0.9795 & 0.9780 & 0.6443 \\ 
    {\our w/o two-pronged} & 0.9747 & 0.9775 & 0.9818 & 0.6485 \\
    {\our w/o adaptive} & 0.9822 & 0.9905 & 0.9802 & 0.6443 \\ 
    {\our w. avgPooling} & 0.9795 & 0.9891 & 0.9798 & 0.6360 \\  
    \hline
    \end{tabular}
  }
  \vspace{-2.5mm}
\end{table}

\begin{table}[!t]
  \centering
  \caption{Average candidate set size and average fraction of truly unionable tables in the set (\%)}   
  \label{tab:candidate}
  \vspace{-3.5mm}
  \resizebox{0.9\linewidth}{!}{
              \setlength{\tabcolsep}{2pt}
    \begin{tabular}{lcccc}
    \toprule
    {}  & \santosSmall & \TUSSmall & \TUSLarge & \wiki \\
    \hline
    $k$         & 10 & 60 & 60 & 40 \\
    {\our}           & {\bf 12.84 ({94.7}\%)} & {\bf 91.57  (89.9\%)}  & {\bf 114.04 (88.2\%)} & {\bf 44.62 (60.2\%)} \\
    {\starmie}       & 41.24 (49.9\%) & 167.91 (68.9\%) & 247.45  (51.3\%) & 69.62 (25.6\%) \\ 
    {\liftus}        & 30.90 (63.0\%) & 222.90 (51.6\%) & 228.45  (63.1\%) & 65.76 (24.8\%) \\ 
    \hline
    \end{tabular}
  }
  \vspace{-3mm}
\end{table}
\header 
\textbf{Study on Candidate Set $\candset$.} \cref{tab:candidate} reports, for each method, the average candidate set size and the fraction of truly unionable tables per candidate set. Our method \our consistently retrieves a much smaller candidate set $\candset$ with a higher fraction of truly unionable tables across all datasets, and the set size is close to $k$. For example, on \santosSmall with $k=10$, \our yields an average candidate set size of 12.84 with 94.7\% truly unionable tables, while \starmie and \liftus return much larger sets of 41.24 and 30.90 candidates with only 49.9\% and 63.0\% truly unionable tables, respectively. Similar trends are observed on other datasets. These results demonstrate the effectiveness of our table-centric adaptive candidate retrieval in \cref{sec:retrieval}, which leverages table embeddings that capture holistic table-level unionability scores $\scOne$ to directly retrieve a compact, high-quality candidate set $\candset$ adaptive to the distribution of $\scOne$ scores.
In contrast, column-centric methods such as \starmie and \liftus include a table in the candidate set if any of its columns matches any query column, while a query table has multiple columns, resulting in larger candidate sets with a lower fraction of truly unionable tables.

\begin{figure}
    \centering
    \includegraphics[width=0.8\linewidth]{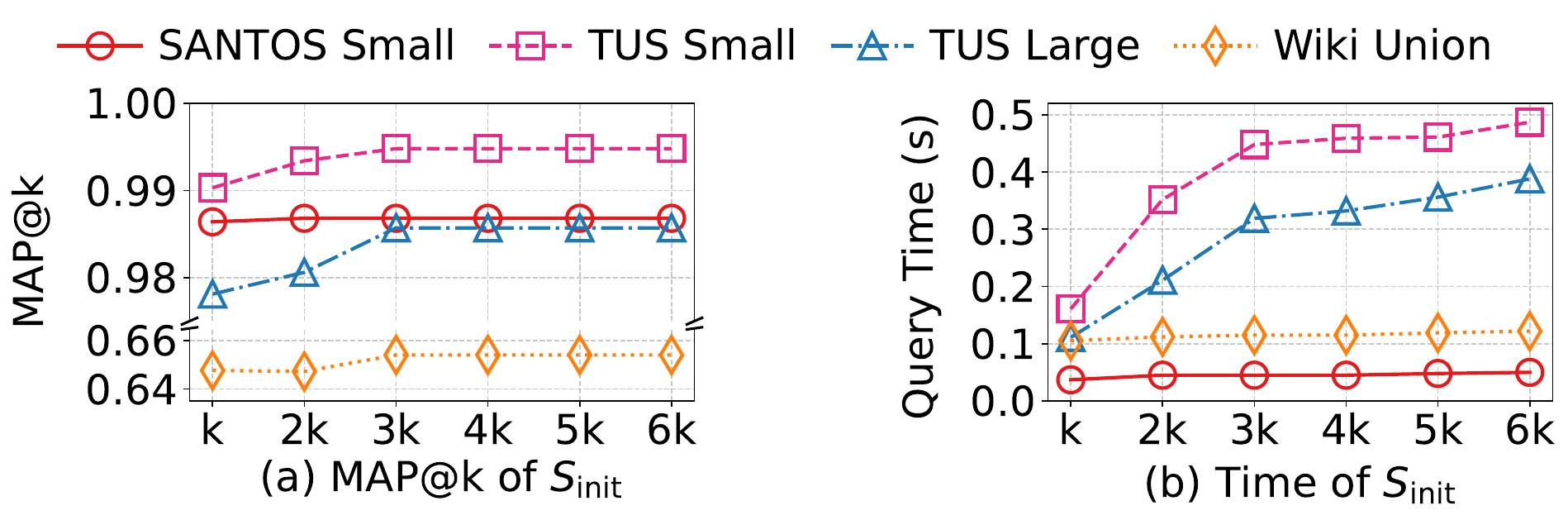}
    \vspace{-1mm}
    \caption{Vary $|\Cinit|$} 
    \label{fig:vary3k} 
    \vspace{-2mm}
\end{figure}

\header 
\textbf{The Effect of the Size of $\Cinit$.} By default, we set the initial candidate set size $|\Cinit|$ to $3k$ in \cref{alg:cand-retrival} (\cref{sec:retrieval}). To access its effect, we vary $|\Cinit|$ from $k$ to $6k$ and report MAP@k and online search time in \cref{fig:vary3k}. As $|\Cinit|$ increases from $k$ to $3k$, MAP@k improves due to the inclusion of more unionable tables, but plateaus for larger values ($4k$ to $6k$), while online search time continues to increase. Thus, setting $|\Cinit|=3k$ is sufficient to achieve high effectiveness with reasonable overhead.

\begin{figure}
    \centering
    \includegraphics[width=\linewidth]{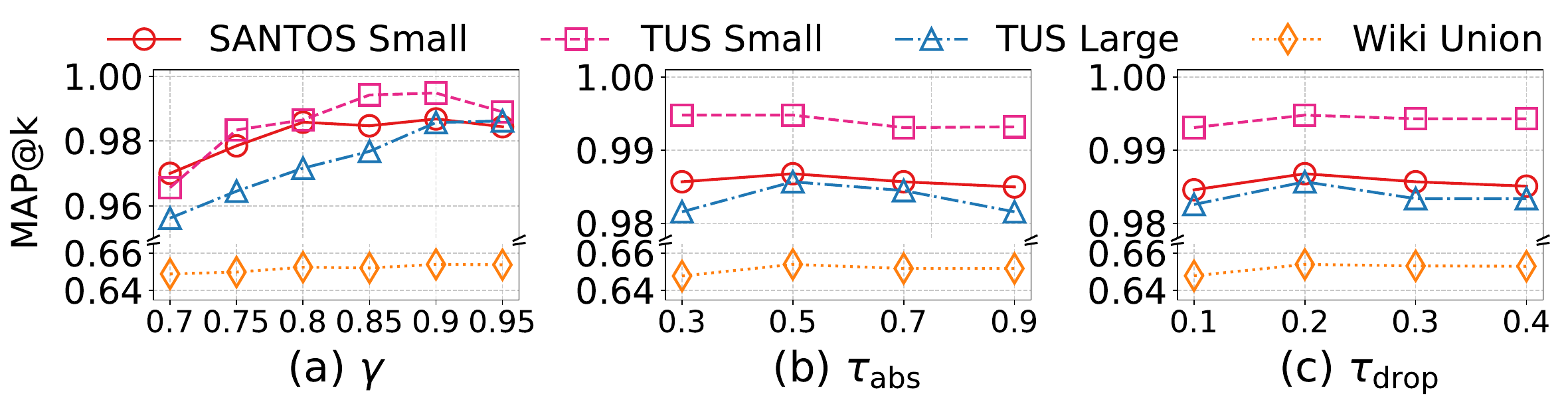}
    \vspace{-4mm}
    \caption{Vary Parameters}
    \label{fig:parameter} 
    \vspace{-2mm}
\end{figure}

\header 
\textbf{Parameter Analysis.} We use the same parameter settings for all datasets by default. Here, we analyze the impact of varying key parameters, including $\gamma$ in \cref{eq:latent-set} for identifying latent positives, and $\tau_{\text{abs}}$ and $\tau_{\text{drop}}$ in \cref{sec:retrieval} for adaptive candidate retrieval. 
In \cref{fig:parameter}(a), as $\gamma$ increases from 0.7 to 0.9, performance improves because more accurate latent positives are excluded from the negative set; further increasing $\gamma$ causes performance to plateau or decrease, as an overly strict threshold may fail to exclude enough latent positives. Thus, we set $\gamma=0.9$ by default.
In \cref{fig:parameter}(b,c), we vary $\tau_{\text{abs}}$ from 0.3 to 0.9 and $\tau_{\text{drop}}$ from 0.1 to 0.4. For both parameters, performance increases and then decreases as the values grow. Therefore, we set $\tau_{\text{abs}}=0.5$ and $\tau_{\text{drop}}=0.2$ by default.

\begin{table}[t]
\centering
\caption{Comparison of FastText and BERT for reranking in accuracy (MAP@k) and encoding time (s)}
\label{tab:fasttext-bert-rerank}
\vspace{-3mm}
\resizebox{\linewidth}{!}{
                \setlength{\tabcolsep}{2pt}
    \begin{tabular}{lcccc}
    \toprule
    \textbf{Model} 
    & \textbf{SANTOS Small} 
    & \textbf{TUS Small} 
    & \textbf{TUS Large} 
    & \textbf{Wiki Union} \\
    & MAP@k / Time(s) 
    & MAP@k / Time(s) 
    & MAP@k / Time(s) 
    & MAP@k / Time(s) \\
    \hline
    FastText 
    & {0.9860} / {13} 
    & 0.9936 / {34} 
    & {0.9844} / {71} 
    & 0.6537 / {211} \\
    BERT 
    & 0.9857 / 35 
    & {0.9938} / 92 
    & 0.9828 / 202 
    & {0.6540} / 635 \\
    \hline
    \end{tabular}
}
\vspace{-2.5mm}
\end{table}

\header 
\textbf{Reranking: FastText vs. BERT.}
In \cref{sec:rerank}, we adopt FastText for dual-evidence reranking due to its efficiency and suitability for our design. To further validate this choice, we compare FastText with BERT for reranking. 
\cref{tab:fasttext-bert-rerank} reports the accuracy and efficiency of our method using FastText versus BERT for reranking. Both achieve nearly identical MAP@k, while FastText is substantially more efficient in encoding.

\begin{figure}
    \centering
    \includegraphics[width=\linewidth]{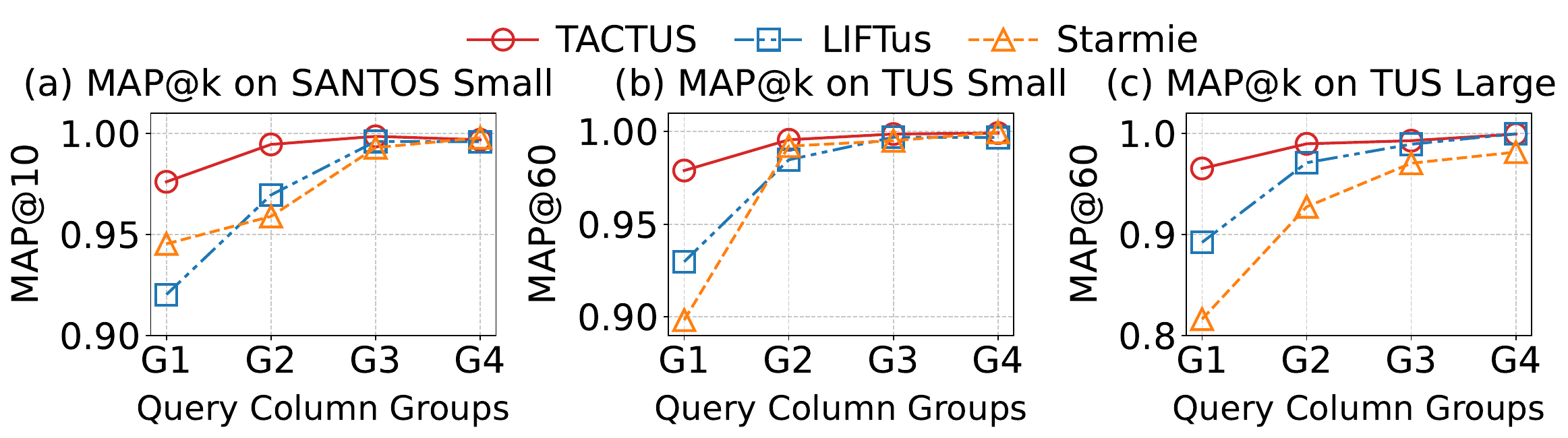}
    \caption{MAP@k by Query Table Width}
    \label{fig:mapk-by-query-quart} 
    \vspace{-3mm}
\end{figure}

\header
\textbf{Vary Query Table Width.} 
We analyze the impact of query table width by sorting all query tables in ascending order of their number of columns and dividing them into four equal-sized groups, denoted as G1 (narrowest) to G4 (widest). 
As shown in \cref{fig:mapk-by-query-quart}, the performance of all methods generally improves as table width increases from G1 to G4, as wider tables provide richer contextual information for unionability search. 
\our consistently outperforms the baselines; for example, on \santosSmall in G2, \our achieves $MAP@k$ of 99.45\%, which is 2.50\% higher than \liftus (96.95\%) and 3.56\% higher than \starmie (95.89\%).

\begin{figure}
    \centering
    \includegraphics[width=0.75\linewidth]{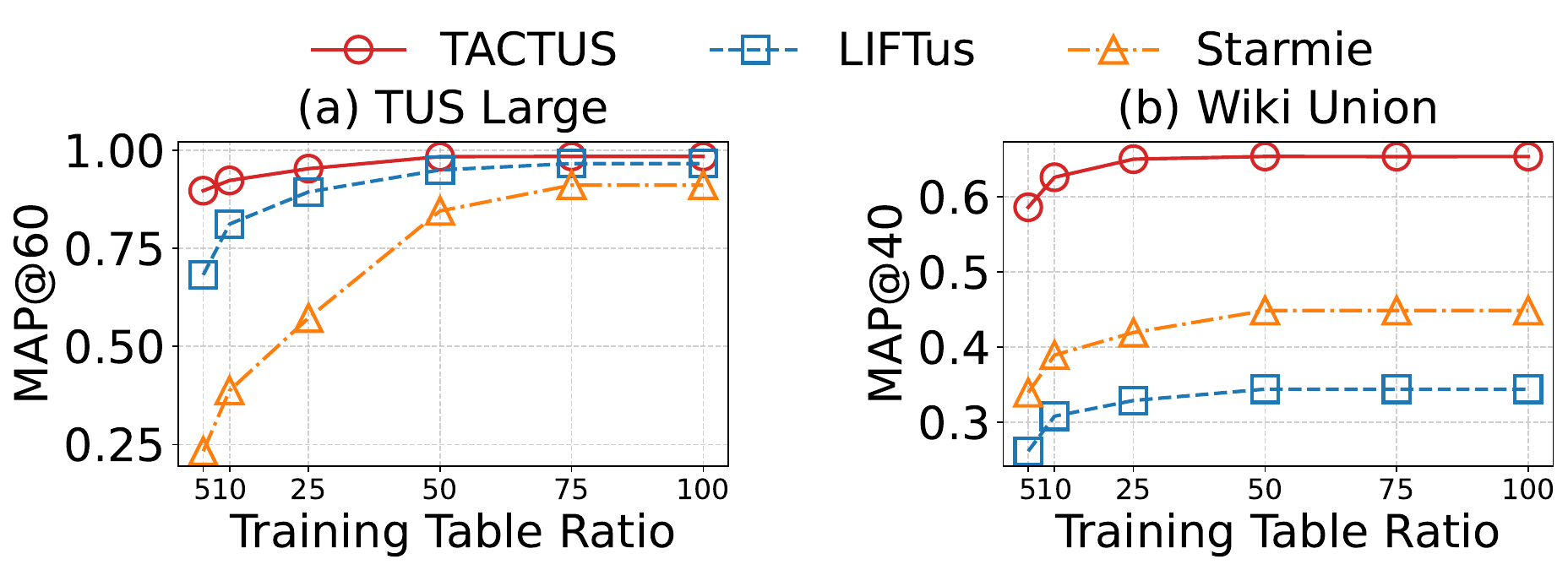}
    \caption{MAP@k when varying training data ratio (\%)}
    \label{fig:training-effi}
        \vspace{-3mm}
\end{figure}

\header\textbf{Training Efficiency.}  
In \cref{fig:training-effi}, we evaluate the training efficiency of \our, \starmie, and \liftus by varying the fraction of training tables used on \TUSLarge and \wiki. As the proportion of training data increases from 5\% to 100\%, all methods improve in $MAP@k$, while \our consistently achieves the highest performance across all fractions. The gains of \our are particularly pronounced in the low-data regime (e.g., 5\% or 10\% of training tables). For example, in \cref{fig:training-effi}(b) on \wiki, when using only 10\% of the training data, \our attains $MAP@k$ of {62.60}\%, which is {31.82\%} higher than \liftus and {23.68\%} higher than \starmie. These results demonstrate that our table-centric method \our is especially effective and data-efficient under limited training data.

\begin{table}[t]
    \centering
    \caption{Embedding Index Size (MB)}
    \label{tab:embedding-store} 
    \vspace{-3.5mm}
    \resizebox{\linewidth}{!}{
          \setlength{\tabcolsep}{2pt}
    \begin{tabular}{lcccccc}
    \toprule
    {Method} & {\santosSmall} & {\santosLarge} & {\TUSSmall} & {
    \TUSLarge} & {\wiki} & {\WDC}\\
    \hline
    \our      & {\bf 10.9}  & {\bf 196.9}  & {\bf 27.4}   & {\bf 97.1}   & {\bf 396.0}   & {\bf 12779.5} \\
    \starmie  & 38.7  & 746.6  & 58.2   & 215.8  & 420.9   & 34570.2 \\
    \liftus   & 519.8 & 7811.9 & 1218.9 & 3719.5 & 10040.9 &  414860.8 \\
    \sato     & 104.6 & 2242.6 & 244.6  & 1013.8 & 1760.2  &  94589.7 \\
    \sherlock & 116.8 & 2252.8 & 276.6  & 1015.2 & 1975.2  & 100526.1\\
    \hline
    \end{tabular}
    }
    \vspace{-2.5mm}
\end{table}

\header 
\textbf{Embedding Index Size.} We report the total size of all embeddings and the index built over these embeddings in \cref{tab:embedding-store}.  \our requires less space than all baselines across all datasets for embedding and index storage. For example, on \TUSLarge, \our requires only 97.1 MB, while \liftus, the runner-up in effectiveness in \cref{tab:overall}, requires 3719.5 MB and \starmie requires 215.8 MB. This demonstrates the space efficiency of our techniques.  

\section{Conclusion}
\label{sec:conclusion}

We present \our, a novel table-centric framework for efficient and accurate table union search in large data lakes. Unlike existing column-centric approaches, \our model holistic table-level unionability through dedicated table embeddings, and enable efficient and effective online search by table-centric adaptive candidate retrieval and dual-evidence reranking techniques. Our  offline processing introduces positive table pair construction, two-pronged negative sampling, and attentive table encoder   for effective table-level unionability estimation. 
Extensive experiments demonstrate that \our consistently achieves state-of-the-art effectiveness and efficiency across diverse datasets. 
This work underscores the value of table-level modeling for tabular data discovery. Future directions include extending our framework to other tasks such as table join search and schema matching, and exploring large language models for table representation learning.



\bibliographystyle{ACM-Reference-Format}
\bibliography{sample}

@String{Computer = "{IEEE} Computer" }

@inproceedings{LiLL08,
  author       = {Chen Li and
                  Jiaheng Lu and
                  Yiming Lu},
  title        = {Efficient Merging and Filtering Algorithms for Approximate String
                  Searches},
  booktitle    = {{ICDE}},
  pages        = {257--266},
  publisher    = {{IEEE} Computer Society},
  year         = {2008}
}

@article{WangXY021,
  author       = {Mengzhao Wang and
                  Xiaoliang Xu and
                  Qiang Yue and
                  Yuxiang Wang},
  title        = {A Comprehensive Survey and Experimental Comparison of Graph-Based
                  Approximate Nearest Neighbor Search},
  journal      = {Proc. {VLDB} Endow.},
  volume       = {14},
  number       = {11},
  pages        = {1964--1978},
  year         = {2021}
}

@article{PanWL24,
  author       = {James Jie Pan and
                  Jianguo Wang and
                  Guoliang Li},
  title        = {Survey of vector database management systems},
  journal      = {{VLDB} J.},
  volume       = {33},
  number       = {5},
  pages        = {1591--1615},
  year         = {2024}
}

@article{LiZSWLZL20,
  author       = {Wen Li and
                  Ying Zhang and
                  Yifang Sun and
                  Wei Wang and
                  Mingjie Li and
                  Wenjie Zhang and
                  Xuemin Lin},
  title        = {Approximate Nearest Neighbor Search on High Dimensional Data - Experiments,
                  Analyses, and Improvement},
  journal      = {{IEEE} Trans. Knowl. Data Eng.},
  volume       = {32},
  number       = {8},
  pages        = {1475--1488},
  year         = {2020}
}

@inproceedings{google_dataset_search,
  author       = {Dan Brickley and
                  Matthew Burgess and
                  Natasha F. Noy},
  title        = {Google Dataset Search: Building a search engine for datasets in an
                  open Web ecosystem},
  booktitle    = {{WWW}},
  pages        = {1365--1375},
  publisher    = {{ACM}},
  year         = {2019}
}

@article{castelo2021auctus,
  author       = {Sonia Castelo and
                  R{\'{e}}mi Rampin and
                  A{\'{e}}cio S. R. Santos and
                  Aline Bessa and
                  Fernando Chirigati and
                  Juliana Freire},
  title        = {Auctus: {A} Dataset Search Engine for Data Discovery and Augmentation},
  journal      = {Proc. {VLDB} Endow.},
  volume       = {14},
  number       = {12},
  pages        = {2791--2794},
  year         = {2021}
}

@inproceedings{fernandez2018aurum,
  author       = {Raul Castro Fernandez and
                  Ziawasch Abedjan and
                  Famien Koko and
                  Gina Yuan and
                  Samuel Madden and
                  Michael Stonebraker},
  title        = {Aurum: {A} Data Discovery System},
  booktitle    = {{ICDE}},
  pages        = {1001--1012},
  year         = {2018}
}

@article{limaye2010annotating,
  author       = {Girija Limaye and
                  Sunita Sarawagi and
                  Soumen Chakrabarti},
  title        = {Annotating and Searching Web Tables Using Entities, Types and Relationships},
  journal      = {Proc. {VLDB} Endow.},
  volume       = {3},
  number       = {1},
  pages        = {1338--1347},
  year         = {2010}
}

@inproceedings{santos2022sketch,
  author       = {A{\'{e}}cio S. R. Santos and
                  Aline Bessa and
                  Christopher Musco and
                  Juliana Freire},
  title        = {A Sketch-based Index for Correlated Dataset Search},
  booktitle    = {{ICDE}},
  pages        = {2928--2941},
  publisher    = {{IEEE}},
  year         = {2022}
}

@article{adelfio2013schema,
  author       = {Marco D. Adelfio and
                  Hanan Samet},
  title        = {Schema Extraction for Tabular Data on the Web},
  journal      = {Proc. {VLDB} Endow.},
  volume       = {6},
  number       = {6},
  pages        = {421--432},
  year         = {2013}
}

@inproceedings{farid2016clams,
  author       = {Mina H. Farid and
                  Alexandra Roatis and
                  Ihab F. Ilyas and
                  Hella{-}Franziska Hoffmann and
                  Xu Chu},
  title        = {{CLAMS:} Bringing Quality to Data Lakes},
  booktitle    = {{SIGMOD} Conference},
  pages        = {2089--2092},
  publisher    = {{ACM}},
  year         = {2016}
}

@article{nargesian2019data,
  author       = {Fatemeh Nargesian and
                  Erkang Zhu and
                  Ren{\'{e}}e J. Miller and
                  Ken Q. Pu and
                  Patricia C. Arocena},
  title        = {Data Lake Management: Challenges and Opportunities},
  journal      = {Proc. {VLDB} Endow.},
  volume       = {12},
  number       = {12},
  pages        = {1986--1989},
  year         = {2019}
}

@article{nargesian2018table,
  author       = {Fatemeh Nargesian and
                  Erkang Zhu and
                  Ken Q. Pu and
                  Ren{\'{e}}e J. Miller},
  title        = {Table Union Search on Open Data},
  journal      = {Proc. {VLDB} Endow.},
  volume       = {11},
  number       = {7},
  pages        = {813--825},
  year         = {2018}
}

@article{miller2018open,
  author       = {Ren{\'{e}}e J. Miller},
  title        = {Open Data Integration},
  journal      = {Proc. {VLDB} Endow.},
  volume       = {11},
  number       = {12},
  pages        = {2130--2139},
  year         = {2018}
}

@article{lehmberg2017stitching,
  author       = {Oliver Lehmberg and
                  Christian Bizer},
  title        = {Stitching Web Tables for Improving Matching Quality},
  journal      = {Proc. {VLDB} Endow.},
  volume       = {10},
  number       = {11},
  pages        = {1502--1513},
  year         = {2017}
}

@inproceedings{d3l,
  author       = {Alex Bogatu and
                  Alvaro A. A. Fernandes and
                  Norman W. Paton and
                  Nikolaos Konstantinou},
  title        = {Dataset Discovery in Data Lakes},
  booktitle    = {{ICDE}},
  pages        = {709--720},
  publisher    = {{IEEE}},
  year         = {2020}
}

@article{santos,
  author       = {Aamod Khatiwada and
                  Grace Fan and
                  Roee Shraga and
                  Zixuan Chen and
                  Wolfgang Gatterbauer and
                  Ren{\'{e}}e J. Miller and
                  Mirek Riedewald},
  title        = {{SANTOS:} Relationship-based Semantic Table Union Search},
  journal      = {Proc. {ACM} Manag. Data},
  volume       = {1},
  number       = {1},
  pages        = {9:1--9:25},
  year         = {2023}
}

@inproceedings{sherlock,
  author       = {Madelon Hulsebos and
                  Kevin Zeng Hu and
                  Michiel A. Bakker and
                  Emanuel Zgraggen and
                  Arvind Satyanarayan and
                  Tim Kraska and
                  {\c{C}}agatay Demiralp and
                  C{\'{e}}sar A. Hidalgo},
  title        = {Sherlock: {A} Deep Learning Approach to Semantic Data Type Detection},
  booktitle    = {{KDD}},
  pages        = {1500--1508},
  publisher    = {{ACM}},
  year         = {2019}
}

@article{sato,
  author       = {Dan Zhang and
                  Yoshihiko Suhara and
                  Jinfeng Li and
                  Madelon Hulsebos and
                  {\c{C}}agatay Demiralp and
                  Wang{-}Chiew Tan},
  title        = {Sato: Contextual Semantic Type Detection in Tables},
  journal      = {Proc. {VLDB} Endow.},
  volume       = {13},
  number       = {11},
  pages        = {1835--1848},
  year         = {2020}
}

@article{starmie,
  author       = {Grace Fan and
                  Jin Wang and
                  Yuliang Li and
                  Dan Zhang and
                  Ren{\'{e}}e J. Miller},
  title        = {Semantics-aware Dataset Discovery from Data Lakes with Contextualized
                  Column-based Representation Learning},
  journal      = {Proc. {VLDB} Endow.},
  volume       = {16},
  number       = {7},
  pages        = {1726--1739},
  year         = {2023}
}

@inproceedings{liftus,
  author       = {Ermu Qiu and
                  Jun Gao and
                  Yaofeng Tu and
                  Jingru Yang},
  title        = {LIFTus: An Adaptive Multi-Aspect Column Representation Learning for
                  Table Union Search},
  booktitle    = {{ICDE}},
  pages        = {2174--2187},
  publisher    = {{IEEE}},
  year         = {2025}
}

@inproceedings{lehmberg2016large,
  author       = {Oliver Lehmberg and
                  Dominique Ritze and
                  Robert Meusel and
                  Christian Bizer},
  title        = {A Large Public Corpus of Web Tables containing Time and Context Metadata},
  booktitle    = {{WWW} (Companion Volume)},
  pages        = {75--76},
  publisher    = {{ACM}},
  year         = {2016}
}

@inproceedings{sarma2012finding,
  author       = {Anish Das Sarma and
                  Lujun Fang and
                  Nitin Gupta and
                  Alon Y. Halevy and
                  Hongrae Lee and
                  Fei Wu and
                  Reynold Xin and
                  Cong Yu},
  title        = {Finding related tables},
  booktitle    = {{SIGMOD} Conference},
  pages        = {817--828},
  publisher    = {{ACM}},
  year         = {2012}
}

@inproceedings{zhang2020finding,
  author       = {Yi Zhang and
                  Zachary G. Ives},
  title        = {Finding Related Tables in Data Lakes for Interactive Data Science},
  booktitle    = {{SIGMOD} Conference},
  pages        = {1951--1966},
  publisher    = {{ACM}},
  year         = {2020}
}

@article{tu2023unicorn,
  author       = {Jianhong Tu and
                  Ju Fan and
                  Nan Tang and
                  Peng Wang and
                  Guoliang Li and
                  Xiaoyong Du and
                  Xiaofeng Jia and
                  Song Gao},
  title        = {Unicorn: {A} Unified Multi-tasking Model for Supporting Matching Tasks
                  in Data Integration},
  journal      = {Proc. {ACM} Manag. Data},
  volume       = {1},
  number       = {1},
  pages        = {84:1--84:26},
  year         = {2023}
}

@article{PimplikarS12,
  author       = {Rakesh Pimplikar and
                  Sunita Sarawagi},
  title        = {Answering Table Queries on the Web using Column Keywords},
  journal      = {Proc. {VLDB} Endow.},
  volume       = {5},
  number       = {10},
  pages        = {908--919},
  year         = {2012}
}

@article{abs-2505-21329,
  author       = {Allaa Boutaleb and
                  Bernd Amann and
                  Hubert Naacke and
                  Rafael Angarita},
  title        = {Something's Fishy In The Data Lake: {A} Critical Re-evaluation
                  of Table Union Search Benchmarks},
  journal      = {CoRR},
  volume       = {abs/2505.21329},
  year         = {2025}
}

@article{cafarella2009data,
  author       = {Michael J. Cafarella and
                  Alon Y. Halevy and
                  Nodira Khoussainova},
  title        = {Data Integration for the Relational Web},
  journal      = {Proc. {VLDB} Endow.},
  volume       = {2},
  number       = {1},
  pages        = {1090--1101},
  year         = {2009}
}

@inproceedings{bert,
  author       = {Jacob Devlin and
                  Ming{-}Wei Chang and
                  Kenton Lee and
                  Kristina Toutanova},
  title        = {{BERT:} Pre-training of Deep Bidirectional Transformers for Language
                  Understanding},
  booktitle    = {{NAACL-HLT} {(1)}},
  pages        = {4171--4186},
  year         = {2019}
}

@inproceedings{simclr,
  author       = {Ting Chen and
                  Simon Kornblith and
                  Mohammad Norouzi and
                  Geoffrey E. Hinton},
  title        = {A Simple Framework for Contrastive Learning of Visual Representations},
  booktitle    = {{ICML}},
  series       = {Proceedings of Machine Learning Research},
  volume       = {119},
  pages        = {1597--1607},
  year         = {2020}
}

@article{hnsw,
  author       = {Yury A. Malkov and
                  Dmitry A. Yashunin},
  title        = {Efficient and Robust Approximate Nearest Neighbor Search Using Hierarchical
                  Navigable Small World Graphs},
  journal      = {{IEEE} Trans. Pattern Anal. Mach. Intell.},
  volume       = {42},
  number       = {4},
  pages        = {824--836},
  year         = {2020}
}

@inproceedings{ling2013synthesizing,
  author       = {Xiao Ling and
                  Alon Y. Halevy and
                  Fei Wu and
                  Cong Yu},
  title        = {Synthesizing Union Tables from the Web},
  booktitle    = {{IJCAI}},
  pages        = {2677--2683},
  year         = {2013}
}

@inproceedings{vaswani2017attention,
  author       = {Ashish Vaswani and
                  Noam Shazeer and
                  Niki Parmar and
                  Jakob Uszkoreit and
                  Llion Jones and
                  Aidan N. Gomez and
                  Lukasz Kaiser and
                  Illia Polosukhin},
  title        = {Attention is All you Need},
  booktitle    = {{NIPS}},
  pages        = {5998--6008},
  year         = {2017}
}

@inproceedings{harmouch2021relational,
  author       = {Hazar Harmouch and
                  Thorsten Papenbrock and
                  Felix Naumann},
  title        = {Relational Header Discovery using Similarity Search in a Table Corpus},
  booktitle    = {{ICDE}},
  pages        = {444--455},
  publisher    = {{IEEE}},
  year         = {2021}
}

@article{PugnaloniZPLNS25,
  author       = {Francesco Pugnaloni and
                  Luca Zecchini and
                  Matteo Paganelli and
                  Matteo Lissandrini and
                  Felix Naumann and
                  Giovanni Simonini},
  title        = {Table Overlap Estimation through Graph Embeddings},
  journal      = {Proc. {ACM} Manag. Data},
  volume       = {3},
  number       = {3},
  pages        = {228:1--228:25},
  year         = {2025}
}

@inproceedings{zhu2019josie,
  author       = {Erkang Zhu and
                  Dong Deng and
                  Fatemeh Nargesian and
                  Ren{\'{e}}e J. Miller},
  title        = {{JOSIE:} Overlap Set Similarity Search for Finding Joinable Tables
                  in Data Lakes},
  booktitle    = {{SIGMOD} Conference},
  pages        = {847--864},
  publisher    = {{ACM}},
  year         = {2019}
}

@inproceedings{dong2021efficient,
  author       = {Yuyang Dong and
                  Kunihiro Takeoka and
                  Chuan Xiao and
                  Masafumi Oyamada},
  title        = {Efficient Joinable Table Discovery in Data Lakes: {A} High-Dimensional
                  Similarity-Based Approach},
  booktitle    = {{ICDE}},
  pages        = {456--467},
  publisher    = {{IEEE}},
  year         = {2021}
}

@article{dong2023deepjoin,
  author       = {Yuyang Dong and
                  Chuan Xiao and
                  Takuma Nozawa and
                  Masafumi Enomoto and
                  Masafumi Oyamada},
  title        = {DeepJoin: Joinable Table Discovery with Pre-trained Language Models},
  journal      = {Proc. {VLDB} Endow.},
  volume       = {16},
  number       = {10},
  pages        = {2458--2470},
  year         = {2023}
}

@inproceedings{cappuzzo2020creating,
  author       = {Riccardo Cappuzzo and
                  Paolo Papotti and
                  Saravanan Thirumuruganathan},
  title        = {Creating Embeddings of Heterogeneous Relational Datasets for Data
                  Integration Tasks},
  booktitle    = {{SIGMOD} Conference},
  pages        = {1335--1349},
  publisher    = {{ACM}},
  year         = {2020}
}

@article{deng2022turl,
  author       = {Xiang Deng and
                  Huan Sun and
                  Alyssa Lees and
                  You Wu and
                  Cong Yu},
  title        = {{TURL:} Table Understanding through Representation Learning},
  journal      = {{SIGMOD} Rec.},
  volume       = {51},
  number       = {1},
  pages        = {33--40},
  year         = {2022}
}

@inproceedings{suhara2022annotating,
  author       = {Yoshihiko Suhara and
                  Jinfeng Li and
                  Yuliang Li and
                  Dan Zhang and
                  {\c{C}}agatay Demiralp and
                  Chen Chen and
                  Wang{-}Chiew Tan},
  title        = {Annotating Columns with Pre-trained Language Models},
  booktitle    = {{SIGMOD} Conference},
  pages        = {1493--1503},
  publisher    = {{ACM}},
  year         = {2022}
}

@article{miao2023watchog,
  author       = {Zhengjie Miao and
                  Jin Wang},
  title        = {Watchog: {A} Light-weight Contrastive Learning based Framework for
                  Column Annotation},
  journal      = {Proc. {ACM} Manag. Data},
  volume       = {1},
  number       = {4},
  pages        = {272:1--272:24},
  year         = {2023}
}

@inproceedings{koutras2021valentine,
  author       = {Christos Koutras and
                  George Siachamis and
                  Andra Ionescu and
                  Kyriakos Psarakis and
                  Jerry Brons and
                  Marios Fragkoulis and
                  Christoph Lofi and
                  Angela Bonifati and
                  Asterios Katsifodimos},
  title        = {Valentine: Evaluating Matching Techniques for Dataset Discovery},
  booktitle    = {{ICDE}},
  pages        = {468--479},
  publisher    = {{IEEE}},
  year         = {2021}
}

@article{bussotti2023generation,
  author       = {Jean{-}Flavien Bussotti and
                  Enzo Veltri and
                  Donatello Santoro and
                  Paolo Papotti},
  title        = {Generation of Training Examples for Tabular Natural Language Inference},
  journal      = {Proc. {ACM} Manag. Data},
  volume       = {1},
  number       = {4},
  pages        = {243:1--243:27},
  year         = {2023}
}

@inproceedings{fasttext,
  author       = {Armand Joulin and
                  Edouard Grave and
                  Piotr Bojanowski and
                  Tom{\'{a}}s Mikolov},
  title        = {Bag of Tricks for Efficient Text Classification},
  booktitle    = {{EACL} {(2)}},
  pages        = {427--431},
  year         = {2017}
}

@article{srinivas2023lakebench,
  author       = {Kavitha Srinivas and
                  Julian Dolby and
                  Ibrahim Abdelaziz and
                  Oktie Hassanzadeh and
                  Harsha Kokel and
                  Aamod Khatiwada and
                  Tejaswini Pedapati and
                  Subhajit Chaudhury and
                  Horst Samulowitz},
  title        = {LakeBench: Benchmarks for Data Discovery over Data Lakes},
  journal      = {CoRR},
  volume       = {abs/2307.04217},
  year         = {2023}
}

@article{vrandevcic2014wikidata,
  author       = {Denny Vrandecic and
                  Markus Kr{\"{o}}tzsch},
  title        = {Wikidata: a free collaborative knowledgebase},
  journal      = {Commun. {ACM}},
  volume       = {57},
  number       = {10},
  pages        = {78--85},
  year         = {2014}
}

@inproceedings{wolf2020transformers,
  author       = {Thomas Wolf and
                  Lysandre Debut and
                  Victor Sanh and
                  Julien Chaumond and
                  Clement Delangue and
                  Anthony Moi and
                  Pierric Cistac and
                  Tim Rault and
                  R{\'{e}}mi Louf and
                  Morgan Funtowicz and
                  Joe Davison and
                  Sam Shleifer and
                  Patrick von Platen and
                  Clara Ma and
                  Yacine Jernite and
                  Julien Plu and
                  Canwen Xu and
                  Teven Le Scao and
                  Sylvain Gugger and
                  Mariama Drame and
                  Quentin Lhoest and
                  Alexander M. Rush},
  title        = {Transformers: State-of-the-Art Natural Language Processing},
  booktitle    = {{EMNLP} (Demos)},
  pages        = {38--45},
  year         = {2020}
}

@inproceedings{chen2023hytrel,
  author       = {Pei Chen and
                  Soumajyoti Sarkar and
                  Leonard Lausen and
                  Balasubramaniam Srinivasan and
                  Sheng Zha and
                  Ruihong Huang and
                  George Karypis},
  title        = {HyTrel: Hypergraph-enhanced Tabular Data Representation Learning},
  booktitle    = {NeurIPS},
  year         = {2023}
}

@article{liu2024magneto,
  author       = {Yurong Liu and
                  Eduardo Pe{\~{n}}a and
                  A{\'{e}}cio S. R. Santos and
                  Eden Wu and
                  Juliana Freire},
  title        = {Magneto: Combining Small and Large Language Models for Schema Matching},
  journal      = {Proc. {VLDB} Endow.},
  volume       = {18},
  number       = {8},
  pages        = {2681--2694},
  year         = {2025}
}

@article{priority-sample,
  author       = {Majid Daliri and
                  Juliana Freire and
                  Christopher Musco and
                  A{\'{e}}cio S. R. Santos and
                  Haoxiang Zhang},
  title        = {Sampling Methods for Inner Product Sketching},
  journal      = {Proc. {VLDB} Endow.},
  volume       = {17},
  number       = {9},
  pages        = {2185--2197},
  year         = {2024}
}

@article{0001LSDT20,
  author       = {Yuliang Li and
                  Jinfeng Li and
                  Yoshihiko Suhara and
                  AnHai Doan and
                  Wang{-}Chiew Tan},
  title        = {Deep Entity Matching with Pre-Trained Language Models},
  journal      = {Proc. {VLDB} Endow.},
  volume       = {14},
  number       = {1},
  pages        = {50--60},
  year         = {2020}
}

@article{LiLSWHT21,
  author       = {Yuliang Li and
                  Jinfeng Li and
                  Yoshihiko Suhara and
                  Jin Wang and
                  Wataru Hirota and
                  Wang{-}Chiew Tan},
  title        = {Deep Entity Matching: Challenges and Opportunities},
  journal      = {{ACM} J. Data Inf. Qual.},
  volume       = {13},
  number       = {1},
  pages        = {1:1--1:17},
  year         = {2021}
}

@inproceedings{robinsoncontrastive,
  author       = {Joshua David Robinson and
                  Ching{-}Yao Chuang and
                  Suvrit Sra and
                  Stefanie Jegelka},
  title        = {Contrastive Learning with Hard Negative Samples},
  booktitle    = {{ICLR}},
  year         = {2021}
}

@inproceedings{IidaTMI21,
  author       = {Hiroshi Iida and
                  Dung Thai and
                  Varun Manjunatha and
                  Mohit Iyyer},
  title        = {{TABBIE:} Pretrained Representations of Tabular Data},
  booktitle    = {{NAACL-HLT}},
  pages        = {3446--3456},
  year         = {2021}
}

@inproceedings{relu,
  author       = {Xavier Glorot and
                  Antoine Bordes and
                  Yoshua Bengio},
  title        = {Deep Sparse Rectifier Neural Networks},
  booktitle    = {{AISTATS}},
  series       = {{JMLR} Proceedings},
  volume       = {15},
  pages        = {315--323},
  year         = {2011}
}

@article{li2023auto,
  author       = {Peng Li and
                  Yeye He and
                  Cong Yan and
                  Yue Wang and
                  Surajit Chaudhuri},
  title        = {Auto-tables: synthesizing multi-step transformations to relationalize
                  tables without using examples},
  journal      = {{VLDB} J.},
  volume       = {34},
  number       = {4},
  pages        = {47},
  year         = {2025}
}

@article{LiHYWC23,
  author       = {Peng Li and
                  Yeye He and
                  Cong Yan and
                  Yue Wang and
                  Surajit Chaudhuri},
  title        = {Auto-Tables: Synthesizing Multi-Step Transformations to Relationalize
                  Tables without Using Examples},
  journal      = {Proc. {VLDB} Endow.},
  volume       = {16},
  number       = {11},
  pages        = {3391--3403},
  year         = {2023}
}

@inproceedings{jia2023getpt,
  author       = {Ran Jia and
                  Haoming Guo and
                  Xiaoyuan Jin and
                  Chao Yan and
                  Lun Du and
                  Xiaojun Ma and
                  Tamara Stankovic and
                  Marko Lozajic and
                  Goran Zoranovic and
                  Igor Ilic and
                  Shi Han and
                  Dongmei Zhang},
  title        = {GetPt: Graph-enhanced General Table Pre-training with Alternate Attention
                  Network},
  booktitle    = {{KDD}},
  pages        = {941--950},
  publisher    = {{ACM}},
  year         = {2023}
}

@article{kayali2024chorus,
  author       = {Moe Kayali and
                  Anton Lykov and
                  Ilias Fountalis and
                  Nikolaos Vasiloglou and
                  Dan Olteanu and
                  Dan Suciu},
  title        = {{CHORUS:} Foundation Models for Unified Data Discovery and Exploration},
  journal      = {Proc. {VLDB} Endow.},
  volume       = {17},
  number       = {8},
  pages        = {2104--2114},
  year         = {2024}
}

@inproceedings{DUST,
  author       = {Aamod Khatiwada and
                  Roee Shraga and
                  Ren{\'{e}}e J. Miller},
  title        = {Diverse Unionable Tuple Search: Novelty-Driven Discovery in Data Lakes},
  booktitle    = {{EDBT}},
  pages        = {42--55},
  publisher    = {OpenProceedings.org},
  year         = {2026}
}

@article{Omnimatch,
  author       = {Christos Koutras and
                  Jiani Zhang and
                  Xiao Qin and
                  Chuan Lei and
                  Vassilis N. Ioannidis and
                  Christos Faloutsos and
                  George Karypis and
                  Asterios Katsifodimos},
  title        = {OmniMatch: Joinability Discovery in Data Products},
  journal      = {Proc. {VLDB} Endow.},
  volume       = {18},
  number       = {11},
  pages        = {4588--4601},
  year         = {2025}
}

@article{Snoopy,
  author       = {Yuxiang Guo and
                  Yuren Mao and
                  Zhonghao Hu and
                  Lu Chen and
                  Yunjun Gao},
  title        = {Snoopy: Effective and Efficient Semantic Join Discovery via Proxy
                  Columns},
  journal      = {{IEEE} Trans. Knowl. Data Eng.},
  volume       = {37},
  number       = {5},
  pages        = {2971--2985},
  year         = {2025}
}

@article{Pneuma,
  author       = {Muhammad Imam Luthfi Balaka and
                  David Alexander and
                  Qiming Wang and
                  Yue Gong and
                  Adila Krisnadhi and
                  Raul Castro Fernandez},
  title        = {Pneuma: Leveraging LLMs for Tabular Data Representation and Retrieval
                  in an End-to-End System},
  journal      = {Proc. {ACM} Manag. Data},
  volume       = {3},
  number       = {3},
  pages        = {200:1--200:28},
  year         = {2025}
}

@article{BIRDIE,
  author       = {Yuxiang Guo and
                  Zhonghao Hu and
                  Yuren Mao and
                  Baihua Zheng and
                  Yunjun Gao and
                  Mingwei Zhou},
  title        = {{BIRDIE:} Natural Language-Driven Table Discovery Using Differentiable
                  Search Index},
  journal      = {Proc. {VLDB} Endow.},
  volume       = {18},
  number       = {7},
  pages        = {2070--2083},
  year         = {2025}
}

@article{Ember,
  author       = {Sahaana Suri and
                  Ihab F. Ilyas and
                  Christopher R{\'{e}} and
                  Theodoros Rekatsinas},
  title        = {Ember: No-Code Context Enrichment via Similarity-Based Keyless Joins},
  journal      = {Proc. {VLDB} Endow.},
  volume       = {15},
  number       = {3},
  pages        = {699--712},
  year         = {2021}
}

@article{MATE,
  author       = {Mahdi Esmailoghli and
                  Jorge{-}Arnulfo Quian{\'{e}}{-}Ruiz and
                  Ziawasch Abedjan},
  title        = {{MATE:} Multi-Attribute Table Extraction},
  journal      = {Proc. {VLDB} Endow.},
  volume       = {15},
  number       = {8},
  pages        = {1684--1696},
  year         = {2022}
}

\end{document}